\documentclass{elsarticle}

\pdfoutput=1
\usepackage{hyperref}
\usepackage{cleveref}

\journal{Computer Meth. Appl. Mech. Engineering}

\usepackage{continuumMechanics}
\newcommand{\cM}{\cTransformedElasticityTensor} 
\newcommand{\cBarM}{\overline{\cM}} 
\newcommand{\tT}{\tensor{T}}
\newcommand{\tCinv}{\tC^{-1}}

\newcommand{\tJ}{\tFthTsr{J}}
\newcommand{\cJ}{\cFthTsr{J}}
\newcommand{\tA}{\tensor{A}}
\newcommand{\cA}{A}
\newcommand{\residualFunction}{\boldsymbol{\mathcal{R}}}
\newcommand{\discreteBody}{\body_d}

\begin{document}

\begin{frontmatter}

\title{A block-coupled Finite Volume methodology for problems of large strain and large displacement}


\author[mymainaddress]{L.R. Azevedo\corref{mycorrespondingauthor}}
\cortext[mycorrespondingauthor]{Corresponding author}
\ead{lrazevedo@protonmail.ch}

\author[cardiffAddress]{P. Cardiff}

\author[rosalesaddress]{F.J. Galindo-Rosales}

\author[mysecondaryaddress]{M. Schafer}

\address[mymainaddress]{Graduate School of Computational Engineering, Technische Universitat Darmstadt, Dolivostrasse 15, 64293 Darmstadt, Germany}
\address[mysecondaryaddress]{Chair of Numerical Methods in Mechanical Engineering, Technische Universitat Darmstadt, Dolivostrasse 15, 64293 Darmstadt, Germany}
\address[rosalesaddress]{Centro de Estudos de Fen\'omenos de Transporte (CEFT), Dept. Engenharia Qu\'imica, Faculdade de Engenharia da Universidade do Porto, 4200-465 Porto, Portugal}
\address[cardiffAddress]{University College Dublin, School of Mechanical and Materials Engineering, Belfield, Ireland}

\begin{abstract}
A nonlinear block-coupled Finite Volume methodology is developed for large displacement and large strain regime. The new methodology uses the same normal and tangential face derivative discretisations found in the original fully coupled cell-centred Finite Volume solution methodology for linear elasticity, meaning that existing block-coupled implementations may easily be extended to include finite strains. Details are given of the novel approach, including use of the Newton-Raphson procedure on a residual functional defined using the linear momentum equation. A number of 2-D benchmark cases have shown that, compared with a segregated procedure, the new approach exhibits errors with many orders of magnitude smaller and a much higher convergence rate.
\end{abstract}


\end{frontmatter}


\section{Introduction}
The Finite Volume Method (FVM) has been been successfully used for computational solid mechanics (CSM) since late 1980s. For a detailed historical review, see e.g. \cite{cardiff2018thirty}. At present, the typically employed formulation is known as Segregated (SEG). This methodology closely resembles the procedures commonly used in fluid dynamics where memory-efficient segregated solution algorithms are used in conjunction with iterative linear solvers. In practice, the linear momentum vector equation is temporarily decoupled into three scalar component equations that are independently solved, where outer Fixed-Point/Picard iterations provide the required coupling \cite{CARDIFF2016100}. It is a flexible method of discretisation, in the sense that it does not constraint the constitutive equations. But, its major drawback is that it can present poor convergence  whenever there is a strong coupling between displacement components \cite{CARDIFF2016100}. To overcome such inadequacy, it was recently proposed by Cardiff et al. \cite{CARDIFF2016100} a block-coupled solution methodology, where inter-component coupling is implicitly included as coefficients in a block matrix; hereafter named BC, it has shown itself to be much faster than SEG for those strongly coupling test cases (by a factor of 2.5-6 times \cite{CARDIFF2016100}). Furthermore, the BC solver resulted in less execution time and memory requirements than a finite element software for the set of cases tested (in fact it was almost 6 times faster and used 8 times less memory). Nevertheless, the current BC formulation is tied to only one constitutive equation and linear elasticity. The current article presents the first attempt to generalize such methodology in order to add support for large strain and large displacement.
The article is constructed as follows: Section \ref{sec:mathematicalModel} outlines the mathematical model, derived from the governing momentum equation and neo-Hookean constitutive relation. The novel nonlinear FV discretisation is presented in Section \ref{sec:numericalMethod}. Subsequently, in Section \ref{sec:methodVerification} it is presented the application of the new approach to five representative benchmark test cases, where accuracy of the method is compared with that of Segregated approach. Finally, the main findings of the current investigation, and suggestions for future works, are given in Section \ref{sec:conclusion}.
\section{Mathematical model}
\label{sec:mathematicalModel}
\ifodd 1{
Neglecting inertia and body forces for clarity, the conservation of linear momentum for an arbitrary body of volume $\subBody$ bounded by surface $\subBodySurface$ with outward facing unit normal $\N$ is given in strong integral form as:
\begin{equation}
	\volumeCellIntegral[\subBody]{\Div{\tFirstPiola}}
	=
	\closedSurfaceCellIntegral[\subBodySurface]{\tFirstPiola \cdot \N}
	= \vv{0}
	\label{eq:integralStepEquationAfterGauss}
\end{equation}
}\else{
The conservation of the linear momentum for an arbitrary body of volume $\subBody$ bounded by surface $\subBodySurface$ with outward facing unit normal $\N$ is given in strong integral form as:
\begin{equation}
	\volumeCellIntegral[\subBody]{\density \ddot{\tU}}
	=
	\closedSurfaceCellIntegral[\subBodySurface]{\firstPiolaInGradU \cdot \N}
	+
	\volumeCellIntegral[\subBody]{\density \vv{b}}
	\label{eq:integralStepEquationAfterGauss}
\end{equation}
where $\density$, $\ddot{\tU}$ and $ \vv{b}$ are the mass density, acceleration and body force, respectively.}\fi The first Piola-Kirchhoff $\tFirstPiola$ is given by 
\begin{equation}
     \tFirstPiola = \tF^T \cdot \tSecondPiola,
\end{equation}
where $\tF$ is the deformation gradient tensor and $\tSecondPiola$ is the second Piola-Kichhoff stress tensor. This work adopts the the compressible and isotropic neo-Hookean hyperelastic model, i.e.
\begin{equation}
    \tSecondPiola
	  	= \mu(\I - \tCinv) + \lambda (\ln J) \tCinv,
\end{equation}
where $\tC = \tF^T \cdot \tF$ is the right Cauchy-Green deformation tensor and $\{\mu, \lambda\}$ are the Lamé constants. The elasticity tensor $\tElasticityTensor = 2 \, \partial \tSecondPiola/\partial \tC$ of this model has the right-minor symmetry (see \ref{elasticBody}).

The discretisation of the new methodology requires a new tensor, say $\tT^d$ $(d = 1,2,3)$, which is a function of another new quantity called \textbf{transformed elasticity tensor} $\tTransformedElasticityTensor = \tF \cdot \tElasticityTensor \underset{(3)}{\cdot} \tF^T$ (the operator $\underset{(3)}{\cdot}$ is a contraction at the third index) and the face normal $\N$, and is defined as
\begin{equation}
  \begin{split}
    \tT^d
    	&= \cM_{a J d L} N_{J} \eBase_a \otimes \eBase_L\\
		&= \cF_{a I} \cElasticityTensor_{I J K L} \cF_{dK} N_{J} \eBase_a \otimes \eBase_L\\
		&= \cF_{a I} \cElasticityTensor_{I J K L} f^d_{K} N_{J} \eBase_a \otimes \eBase_L \quad\quad (\vv{f}^d = f^d_{K} \eBase_K = F_{dK} \eBase_K)\\
		&\equiv (\tF \cdot \tElasticityTensor) \underset{(2,3)}{:} (\N \otimes \vv{f}^d).
  \end{split}
  \label{eq:tdDefinition}.
\end{equation}
Considering the neo-Hookean model, $\tT^d$ is:
\begin{equation}
  \begin{split}
    \tT^d
         = \lambda (\tensor{A} \cdot \N) \otimes (\tC^{-1} \cdot \vv{f}^d)
         + (\mu - \lambda \ln J)
         \Big{[} 
           ( \tensor{A} \cdot \vv{f}^d ) \otimes \vv{b}
          + (\vv{b} \cdot \vv{f}^d ) \tensor{A}
         \Big{]},
  \end{split}
\end{equation}
where $\tensor{A} = \tF \cdot \tC^{-1}$ and $\vv{b} = \N \cdot \tC^{-1} = \tC^{-1} \cdot \N$.

\section{Numerical method}
\label{sec:numericalMethod}
The mathematical model presented in the preceding section is now discretised using a semi-implicit coupled manner and a cell-centred-based FV approach, providing a discrete approximation of the previously presented exact integral. The discretisation procedure is separated into two distinct parts: discretisation of the solution domain and discretisation of the governing equations. If the temporal effects were considered, time would also be discretised into a finite number of time increments, where the mathematical model is solved in a time-marching manner.
\subsection{Solution domain discretisation}
The starting point for a FV discretisation is to decompose the solution spatial domain $\body$, which is usually approximated by arbitrary and finite number $n_C$ of contiguous convex polyhedral cells (also known as finite volume) $\subBody_C$'s bounded by faces that do not overlap. But this work adopts a specific polyhedral: the rectangular cuboid. The reason for choosing rectangular cuboids is to avoid non-conformal (skewed and/or non-conjunctional) mesh \cite{moukalled2015} and the complexities that arise from it. This way, investigation efforts focus only on the “core” (i.e. minimal structure to be fully usable) of the NLBC methodology. Non-essential extensions can be added to NLBC after an extensive investigation of the core.

\begin{figure}
	\begin{center}
	    \includegraphics[width=0.4\paperwidth]{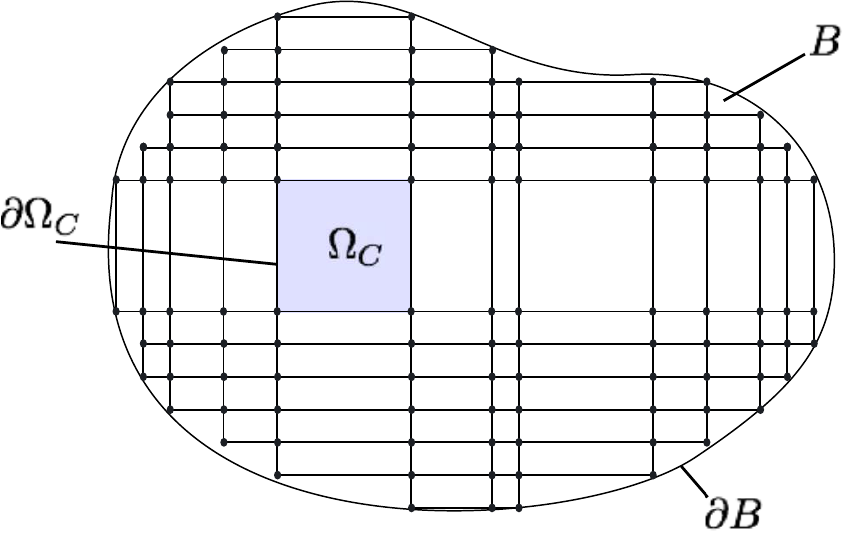}
	    \caption{Discretisation of a body $\body$ into cells $\subBody_C$'s. Every cell $\subBody_C$ has a boundary $\subBodySurface_C$. Note that because of the rectangular cuboid restriction, the boundary domain $\bodySurface$ is approximated in a castellated \textit{staircase} manner.}
	   	\label{fig:discretisationOfBody}
    \end{center}
\end{figure}

\begin{figure}
	\begin{center}
	    \includegraphics[width=1.0\linewidth]{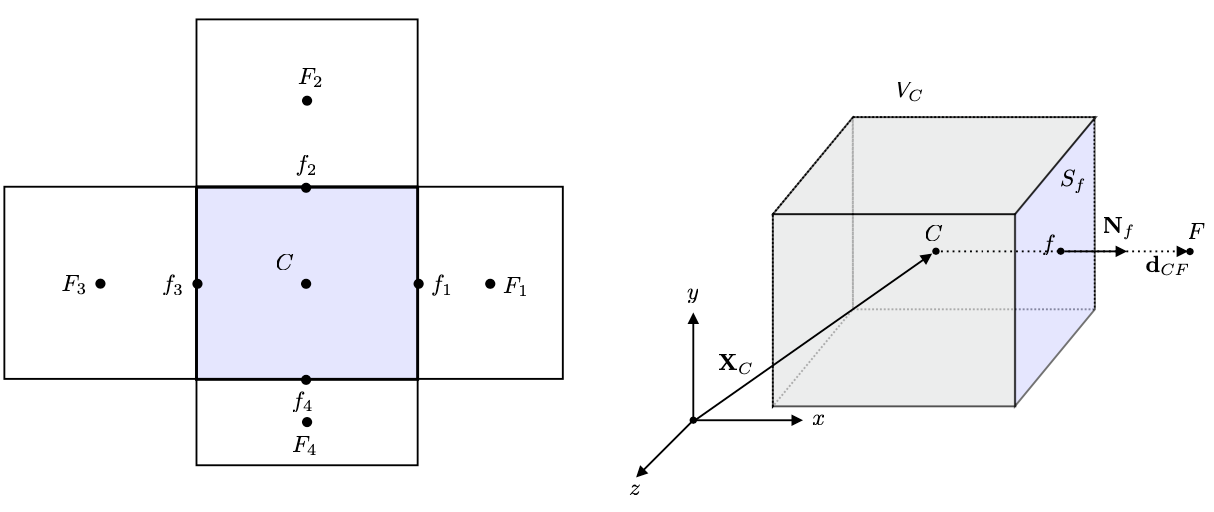}
	    \caption{a) A cell $\subBody_C$, with centroid $C$ of a 2D discretised domain, and its neighbours $F_i$; b) A cell $\subBody_C$ with its geometric parameters used in the finite volume discretisation. In style of \cite{oliveira2017}.}
    	\label{fig:cuboid}
    \end{center}
\end{figure}

The approximation mentioned above is written as
\begin{equation}
	\body
	\approx \discreteBody
	= \bigcup\limits_{C=1}^{n_C} \subBody_C,
\end{equation}
i.e. the continuous body $\body$ is approximated by the computational domain $\discreteBody$ which is the union of $n_C$ cells. The Figure \ref{fig:discretisationOfBody} shows (for two-dimensional case) the configuration of one cell $\subBody_C \in \discreteBody$.
Before proceeding, note the geometric parameters shown in the Figures (\ref{fig:cuboid}a-b) which are needed in the FV discretisation process of the governing equations. The Figure (\ref{fig:cuboid}a) shows a cell with its neighbours $F$'s and their face centroids $f$'s. The other image (Fig. \ref{fig:cuboid}b) exemplifies a typical cuboid cell $\subBody_C$, having volume $V_C$ and the centroid, or computational node, located at the point C.
\subsection{Equation discretisation}
\subsubsection{Momentum equation linearisation}
To solve the governing equation (Eq. (\ref{eq:integralStepEquationAfterGauss})), it is first rewritten as
\begin{equation}
	\residualFunction(\GradU)
	=
	\closedSurfaceCellIntegral[\subBodySurface]
		{
			\firstPiolaInGradU(\GradU) \cdot \N
		}
	= \vv{0}
	\label{eq:momentumEquationToSolve}
\end{equation}
where $\residualFunction$ can be called \textbf{residual function}, since $\residualFunction(\GradU)$ is the so-called, in Finite Element Analysis terminology, residual or out-of-balance force \cite{bonet2008}. The solution of this equation is sought using the Newton-Raphson iterative process whereby, given a solution estimate $\GradU^{n-1}$ at iteration $n-1$, a new value 
$\GradU^{n} = \GradU^{n-1} + \InterGrad^{\circ} \tDeltaU \cdot \tF^{\circ}$ is obtained by establishing the linear approximation:
\begin{equation}
	\begin{split}
	  \residualFunction(\GradU^{n})
	  \approx
	  	\residualFunction(\GradU^{n-1})
	  	+ \FstDev[\residualFunction(\GradU^{n-1})]{\GradU}
	  		: (\InterGrad^{\circ} \tDeltaU \cdot \tF^{\circ})
	  = \vv{0}
	\end{split}
	\label{eq:linearisedMomentumEquationToSolveUsingNnotation}
\end{equation}
where $\InterGrad^{\circ} \tDeltaU$ is the incremental displacement gradient and $\tF^{\circ} = \tF^{n-1}$ (see \ref{incrDrescription}). Using a simplified notation, the equation to be solved is
\begin{equation}
	\begin{split}
	  	\FstDev[\residualFunction(\GradU^{\circ})]{\GradU}
	  		: (\InterGrad^{\circ} \tDeltaU \cdot \tF^{\circ})
	  = - \residualFunction(\GradU^{\circ})
	\end{split}
	\label{eq:linearisedMomentumEquationToSolveUsingOldInArgsNotation}
\end{equation}
which corresponds to
\begin{equation}
	\begin{split}
	    \underbrace{
    	    \closedSurfaceCellIntegral[\subBodySurface]
    		{
    			\Bigg[
    				\FstDev[\firstPiolaInGradU^{\circ}]{\GradU}
    				: (\InterGrad^{\circ} \tDeltaU \cdot \tF^{\circ})
    			\Bigg]
    			 \cdot \N
    		}
		}_{\text{surface force increment}}
		=
		-\underbrace{
    	    \closedSurfaceCellIntegral[\subBodySurface]
    		{
    			\firstPiolaInGradU^{\circ} \cdot \N
    		}
    	}_{\text{old surface force}}
	\end{split}
	\label{eq:system}
\end{equation}
where the simplified notation $\firstPiolaInGradU^{\circ} = \firstPiolaInGradU(\GradU^{\circ})$ has been used.
\subsubsection{Surface force increment term}
\ifodd 3{
The integral of the surface force increment term is approximated as:
\begin{equation}
	\begin{split}{}
		\closedSurfaceCellIntegral[\subBodySurface_C]
		{
			\Bigg[
				\FstDev[\firstPiolaInGradU^{\circ}]{\GradU}
				&: (\InterGrad^{\circ} \tDeltaU \cdot \tF^{\circ})
			\Bigg]
			 \cdot \N
		} \\
		&=
		\sum\displaylimits_{
			\surface_f \in \, \subBodySurface_C
		}
		\surfaceCellIntegral[\surface_f]
		{
			\Bigg[
				\FstDev[\firstPiolaInGradU^{\circ}]{\GradU}
				: (\InterGrad^{\circ} \tDeltaU \cdot \tF^{\circ})
			\Bigg]
			 \cdot \N
		} \quad \quad \text{($\subBodySurface_C$ is a polyhedral)} \\
		&\approx
		\sum\displaylimits_{
			f
		}
		\cSf_f
		\Bigg{[}
			\Bigg{\{}
				\FstDev[\firstPiolaInGradU^{\circ}]{\GradU}
					: (\InterGrad^{\circ} \tDeltaU \cdot \tF^{\circ})
			\Bigg{\}} \cdot \N
		\Bigg{]}_{f}
		 \quad \quad \text{(mid-point rule integration)} .
	\end{split}
	\label{eq:surfaceForceTermDiscretisationFstStep}
\end{equation}
Substituting for $\tA = \InterGrad^{\circ} \tDeltaU \cdot \tF^{\circ}$ into equation (see \ref{elasticBody} for derivation)
\begin{equation}
	\begin{split}
		\FstDev[\firstPiolaInGradU]{\GradU} : \tA
		=
			\tA \cdot \tSecondPiolaInGradU
			+
			\tTransformedElasticityTensor : \tA,
			\quad\quad \forall \tA \in \euclideanVectorSpace^2,
	\end{split}
\end{equation}
yields
\begin{equation}
	\begin{split}
		\Bigg{[}
			\Bigg{\{}
				\FstDev[\firstPiolaInGradU^{\circ}]{\GradU}
					: (\InterGrad^{\circ} \tDeltaU \cdot \tF^{\circ})
			\Bigg{\}} \cdot \N
		\Bigg{]}_{f}
		&= 
		\bigg{[}
			\bigg{(}
				\GradInc \cdot \tSecondPiolaInGradU^{\circ}
			\bigg{)} \cdot \N
		\bigg{]}_{f} \\
		&\quad \quad +
		\bigg{[}
			\bigg{\{}
				\tTransformedElasticityTensor^{\circ}
				: \bigg{(} \GradInc \bigg{)}
			\bigg{\}} \cdot \N
		\bigg{]}_{f},
		\label{eq:firstPiolaInGradIncrement}
	\end{split}
\end{equation}
where $\tSecondPiolaInGradU^{\circ}=\tSecondPiolaInGradU(\GradOld)$ and $\tTransformedElasticityTensor^{\circ} = \tTransformedElasticityTensor(\GradOld)$. Letting $\N\N = \N \otimes \N$, the first term on the right-hand side of Equation (\ref{eq:firstPiolaInGradIncrement}) is
\begin{equation}
	\begin{split}
	\bigg{[}
		\GradInc \cdot \tSecondPiolaInGradU^{\circ} \cdot \N
	\bigg{]}_{f}
	  & =
	  \bigg{[} 
	    (\cGradDU)_{ab} \cF_{bc}^{\circ} \Sigma_{cd}^{\circ} N_d \eBase_a
	  \bigg{]}_{f} \\
	    & =
	  \bigg{[}
	    (\cGradDU)_{ab} v_{b}^{\circ} \eBase_a 
	  \bigg{]}_{f} \quad \quad (\vv{v}^{\circ} = v_{b}^{\circ} \eBase_b = \cF_{bc}^{\circ} \Sigma_{cd}^{\circ} N_d \eBase_b)  \\
	  & =
	  \bigg{[}
	    \tGradDU \cdot \vv{v}^{\circ}
	  \bigg{]}_{f} \\
	  & = 
	  \bigg{[}
	    \tGradDU \cdot \parV{v}^{\circ} + \tGradDU \cdot \othV{v}^{\circ}
	  \bigg{]}_{f}
	      \quad (
	        \parV{v}^{\circ} = (\vv{v}^{\circ} \cdot \N) \N, \,
	        \othV{v}^{\circ} = (\I - \N\N) \cdot \vv{v}^{\circ}
	        ) \\
	  & = 
	  \bigg{[}
	    (\vv{v}^{\circ} \cdot \N)\tGradDU \cdot \N + \tGradDU \cdot \othV{v}^{\circ}
	  \bigg{]}_{f}.
	\end{split}
	\label{eq:firstTermMainEq}
\end{equation}
Note the projection of $\vv{v}^{\circ}$ onto the face normal direction and onto the face plane. This step creates the opportunity to apply the same discretisation procedures, employed by the BC method, to calculate the normal and tangential derivative terms (i.e. $\tGradDU \cdot \N$ and $\tGradDU \cdot \othV{v}^{\circ}$, respectively). The second term on the right-hand side of Equation (\ref{eq:firstPiolaInGradIncrement}) is computed as:
\begin{equation}
	\begin{split}
		\bigg{[}
			\bigg{\{}
				&\tTransformedElasticityTensor^{\circ}
				: \bigg{(} \GradInc \bigg{)}
			\bigg{\}} \cdot \N
		\bigg{]}_{f} \\
		&=
		\bigg{[}
			\cTransformedElasticityTensor_{abcd}^{\circ}
			(\cGradDU)_{ce} \cF_{ed}^{\circ} N_b \eBase_a
		\bigg{]}_{f} \\
		& =
		\bigg{[}
			\cBarM^{\circ}_{acd} (\cGradDU)_{ce} \cF_{ed}^{\circ} \eBase_a
		\bigg{]}_{f} \quad \quad (\cBarM^{\circ}_{acd} = \cM_{abcd}^{\circ} N_b) \\
		& =
		\bigg{[}
			\sum_d \cBarM^{\circ}_{acd} (\cGradDU)_{ce} g_{ed}^{\circ} \eBase_a
		\bigg{]}_{f} \quad \quad 
			(
			\vv{g}_d^{\circ} = g_{ed}^{\circ} \eBase_e = \cF_{ed}^{\circ} \eBase_e
			) \\
		& =
		\bigg{[}
			\sum_d \tT_d^{\circ} \cdot \tGradDU \cdot \vv{g}_d^{\circ}
		\bigg{]}_{f}
			\quad \quad 
			(
			\tT_d^{\circ} = \cBarM^{\circ}_{acd} \eBase_a \otimes \eBase_c
			) \\
		& =
		\bigg{[}
			\sum_d \tT_d^{\circ} \cdot \tGradDU
			\cdot ((\vv{g}_d^{\circ} \cdot \N) \N + (\I - \N\N) \cdot \vv{g}_d^{\circ}))
		\bigg{]}_{f}
		\quad \quad (\text{project } \vv{g}_d^{\circ}) \\
		& =
		\bigg{[}
			\sum_d (\vv{g}_d^{\circ} \cdot \N) \tT_d^{\circ} \cdot (\tGradDU \cdot \N)
		\bigg{]}_{f}
		+
		\bigg{[}
			\sum_d \tT_d^{\circ} \cdot (\tGradDU \cdot \vv{h}_d^{\circ} )
		\bigg{]}_{f} \quad (\vv{h}_d^{\circ} = (\I - \N\N) \cdot \vv{g}_d^{\circ}).
	\end{split}
	\label{eq:secondTermMainEq}
\end{equation}
Using (\ref{eq:firstTermMainEq}) and (\ref{eq:secondTermMainEq}), the term (\ref{eq:firstPiolaInGradIncrement}) is given as:
\begin{equation}
	\begin{split}
			\bigg{[}
				\bigg{\{}
					\FstDev[\firstPiolaInGradU^{\circ}]{\GradU}
					: \left( \GradInc \right)
				\bigg{\}} \cdot \N
			\bigg{]}_{f}
			&=
				\bigg{[}
		    		(\vv{v}^{\circ} \cdot \N)\tGradDU \cdot \N
		    		+ \tGradDU \cdot \othV{v}^{\circ}
		  		\bigg{]}_{f} \\
		  		&+ 
		  		\bigg{[}
					\sum_d (\vv{g}_d^{\circ} \cdot \N ) \tT_d^{\circ}
					\cdot (\tGradDU \cdot \N)
				\bigg{]}_{f} \\
				&+
				\bigg{[}
					\sum_d \tT_d^{\circ}
					\cdot (\tGradDU \cdot \vv{h}_d^{\circ} )
				\bigg{]}_{f} \\
			&=
				\underbrace{\bigg{[}
					\Big{\{}
						(\vv{v}^{\circ} \cdot \N)\I
						+ \sum_d (\vv{g}_d^{\circ} \cdot \N ) \tT_d^{\circ}
					\Big{\}}
					\cdot(\tGradDU \cdot \N) 
					\bigg{]}_{f}}_{(1)} \\
				&+
				\underbrace{\bigg{[}
					\tGradDU \cdot \othV{v}^{\circ}
					+ \sum_d \tT_d^{\circ}
					  \cdot (\tGradDU \cdot \vv{h}_d^{\circ} )
					\bigg{]}_{f}}_{(2)}.
	\end{split}
	\label{eq:firstPiolaInGradIncrementRepeated}
\end{equation}
The underlined terms (1) and (2) from the equation before are approximated using the same approach employed by the BC method \cite{CARDIFF2016100}, i.e. the normal derivative term (1) is discretised using the central differencing method as
\begin{equation}
	\bigg{[}
		\underbrace{
			\Big{\{}
				(\vv{v}^{\circ} \cdot \N)\I
				+ \sum_d (\vv{g}_d^{\circ} \cdot \N ) \tT_d^{\circ}
			\Big{\}}
		}_{\tensor{H}^{\circ}_{n}}
		\cdot \left(\frac{\tDeltaU^C - \tDeltaU^F}{| \vv{d}_{CF} |} \right) 
	\bigg{]}_{f},
	\label{eq:H_nOldTerm}
\end{equation}
where the vector connecting the centroids of the cells sharing the common face $\vv{d}_{CF} = \vv{X}_{F} - \vv{X}_{C}$, and the tangential face derivative term (2) above is discretised using the face-Gauss/Finite Area method as
\begin{equation}
	\begin{split}
	  \bigg{[}
	    \frac{1}{\cSf}
	    \sum_e L_e (&\vv{M}_e \cdot \othV{v}^{\circ}) \tDeltaU_{e}
	    + \sum_d \tT_d^{\circ} \cdot
	    \Big{\{}
	      \frac{1}{| \Sf |} 
	      \sum_e L_e (\vv{M}_e \cdot \vv{h}_d^{\circ}) \tDeltaU_{e} )
	    \Big{\}}
	  \bigg{]}_{f} \\
	  &=
	  \Bigg{[}
	  	\frac{1}{\cSf}
		\sum_e
		\underbrace{
		L_e
		\bigg{[}
		(\vv{M}_e \cdot \othV{v}^{\circ}) \I
		+ \sum_d (\vv{M}_e \cdot \vv{h}_d^{\circ}) \tT_d^{\circ}
		\bigg{]}
		}_{\vv{H}^{\circ}_{t}} \cdot \tDeltaU_{e}
	  \Bigg{]}_{f}.
	\end{split}
	\label{eq:H_tOldTerm}
\end{equation}
\subsubsection{Old surface force}
The discretisation of this term uses the mid-point integration approximation as
\begin{equation}
	\begin{split}
		\closedSurfaceCellIntegral[\subBodySurface_C]
		{
			\firstPiolaInGradU^{\circ} \cdot \N 
		}
		&=
		\sum\displaylimits_{
			\surface_f \in \, \subBodySurface_C
		}
		\surfaceCellIntegral[\surface_f]
		{
			\firstPiolaInGradU^{\circ} \cdot \N 
		} \quad \quad \text{($\subBodySurface_C$ is a polyhedral)} \\
		&\approx 
			\sum\displaylimits_{
					f
				}
			\bigg{[}
				\firstPiolaInGradU^{\circ} \cdot \Sf
			\bigg{]}_{f}	\quad \quad \text{(mid-point rule integration)},
	\end{split}
	\label{eq:oldSurfaceForce}
\end{equation}
where $\firstPiolaInGradU^{\circ}$ is the last known value of the first Piola-Kirchhof stress tensor.
}\else{
The surface integral may be replaced by a sum over the faces of a control volume. The surface force increment term on face $f$ is discretised as follows:
\begin{equation}
	\begin{split}
	    \closedSurfaceCellIntegral[\subBodySurface]
		{
			\Bigg[
				\FstDev[\firstPiolaInGradU(\GradU^{\circ})]{\GradU}
				: (\InterGrad^{\circ} \tDeltaU \cdot \tF^{\circ})
			\Bigg]
			 \cdot \N
		}
		&\approx
	    \sum\displaylimits_{
			f
		}
		\cSf_f
		\Bigg{[}
			\Bigg{\{}
				\FstDev[\firstPiolaInGradU^{\circ}]{\GradU}
					: (\InterGrad^{\circ} \tDeltaU \cdot \tF^{\circ})
			\Bigg{\}} \cdot \N
		\Bigg{]}_{f}\\
		&=
		\sum\displaylimits_{
			f
		}
		\cSf_f
		\bigg{[}
			\bigg{(}
				\GradInc \cdot \tSecondPiolaInGradU^{\circ}
			\bigg{)} \cdot \N
		\bigg{]}_{f} \\
		&\quad \quad +
		\sum\displaylimits_{
			f
		}
		\cSf_f
		\bigg{[}
			\bigg{\{}
				\tTransformedElasticityTensor^{\circ}
				: \bigg{(} \GradInc \bigg{)}
			\bigg{\}} \cdot \N
		\bigg{]}_{f},
	\end{split}
	\label{eq:system}
\end{equation}
}\fi
\subsubsection{Boundary conditions}
The boundary conditions are handled in the same way as in the BC method, except by the fact that, instead of $\tU_f$, $\tU^{\circ}_f + \tDeltaU_f$ is used. Thus, the original Equation $\I\cdot \tU = \tU_b$ in BC, for example, becomes
\begin{equation}
	\begin{split}
		\I \cdot \left( \tU^{\circ}_f + \tDeltaU_f \right)
		= \tU_b
		\implies
		\I \cdot \tDeltaU_f = \tU_b - \tU^{\circ}_f,
	\end{split}
\end{equation}
and Equation (18) in BC becomes
\begin{equation}
	\begin{split}
		\vv{T}_b
		=
		\bigg{[}
			\firstPiolaInGradU(\GradOld) \cdot \N
			+
			\bigg{\{}
				\FstDev[\firstPiolaInGradU(\GradOld)]{\GradU}
				: \bigg{(} \GradInc \bigg{)}
			\bigg{\}} \cdot \N
		\bigg{]}_{f}
	\end{split}
\end{equation}
to be discretised using the same processes applied to the surface force increment and to the old surface force terms described before. The symmetry plane boundary condition is discretised analogously to BC's approach.

\subsection{Solution procedure}
Assembling Equation (\ref{eq:system}) using (\ref{eq:H_nOldTerm}), (\ref{eq:H_tOldTerm}) and (\ref{eq:oldSurfaceForce}) along with the neo-Hookean material model equations produces a linear algebraic equation with the same structure seen in the linear Block-Coupled method. However, instead of solving for the total displacement $\tU$, it is solved for incremental displacement $\tDeltaU$. That is, for each control volume $C$, the final discretised form of the momentum equation can be arranged in the form of $N_i$ linear algebraic equations:
\begin{equation}
	\vv{A}_{C} \cdot \tDeltaU_{C}
	+ \sum\displaylimits_{F} \vv{A}_{F} \cdot \tDeltaU_{F}
	= \vv{R}_{C}
	\label{eq:linearSystemBC}
\end{equation}
where the summation is over the control volume faces. The boundary discretisation creates an additional $N_b$ linear equations with the same structure as Eq. (\ref{eq:linearSystemBC}), one for each boundary face centre. These two sets of linear equations are then assembled forming a linear system of equations:
\begin{equation}
    [\tensor{A}][\tDeltaU]=[\vv{R}]
\end{equation}
where $[\tensor{A}]$ is a sparse $N \times N$ matrix with the tensorial coefficients $\vv{A}_{C}$ on the diagonal and the tensorial coefficients $\vv{A}_{F}$ form the matrix off-diagonal. The total number of computational points being $N = N_i + N_b$.

Just as in BC, the tangential derivative terms contribute solely to the off-diagonal coefficients, thus $[\tensor{A}]$ is not diagonally dominant, in contrast to the segregated methodology. Therefore, the standard preconditioned Conjugate Gradient (CG) methods may not guarantee convergence. As an alternative, the system of linear equations can be solved using, e.g. Bi-Conjugate Gradient Stabilised (BiCGStab), Generalised Minimal Residual (GMRes) or even direct methods \cite{CARDIFF2016100}.

\section{Method verification}
\label{sec:methodVerification}
In this section, the accuracy and robustness of the novel nonlinear block-coupled methodology is examined for five separate representative test cases and comparing the numerical prediction to the available analytical solutions. The methods SEG, BC and NLBC were implemented as a Matlab toolbox called nFVM to generate the results presented in this section. Note that, for all test cases examined here, a solution is considered converged when the residual falls below $10^{-7}$.

\subsection{Infinitesimal elasticity}
It can easily be shown that when the NLBC method is restricted to the linearised elasticity framework, it reduces to BC formulation. Thus, the latter can be seen as a special case of the former. The results from the next test case, that of a slender 2-D cantilever undergoing bending, show this fact by means of numerical simulation. This case was used by Cardiff et al. \cite{CARDIFF2016100} in their seminal work on the BC method.

\subsubsection{Slender cantilever in bending}
\begin{figure}
	\begin{center}
	    \includegraphics[width=0.425\paperwidth]
	    {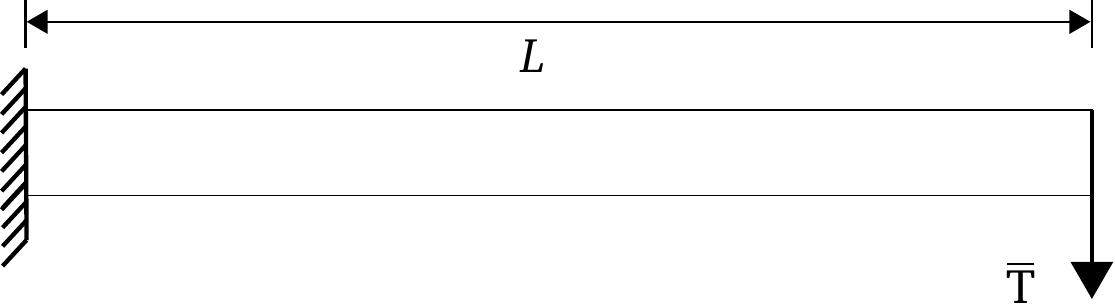}
	    \caption{Geometry and boundary conditions for the slender cantilever beam in bending test case.}
	    \label{fig:slenderCantileverBeam}
    \end{center}
\end{figure}

\begin{figure}
  \begin{center}
    \includegraphics[width=\linewidth]
    {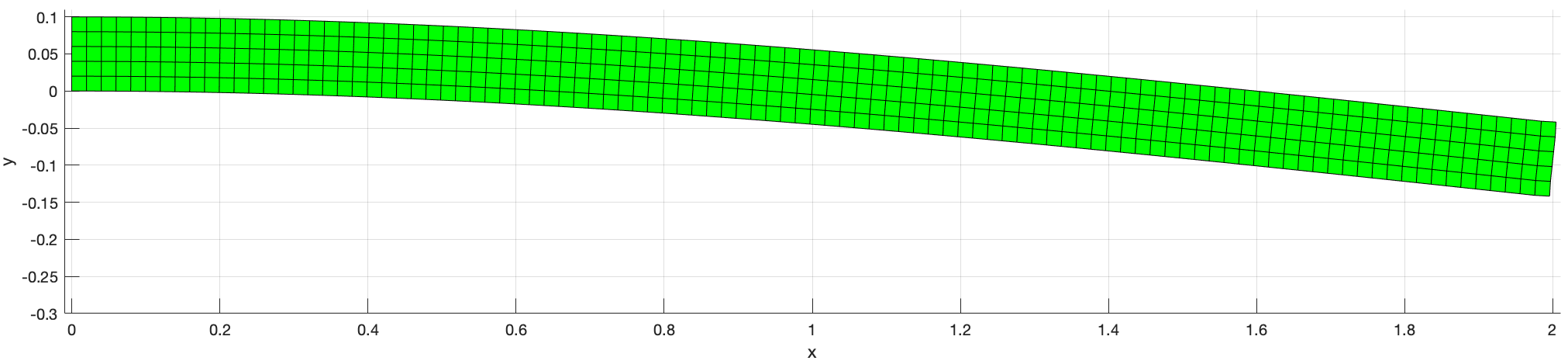}
    \caption{Deformed profile (scaled by factor of 10) for mesh 100x5 cells.}
    \label{fig:scaledCantileverProfile}
  \end{center}
\end{figure}

The geometry of the test case, shown in Figure (\ref{fig:slenderCantileverBeam}), consists of a rectangle beam 2 x 0.1 m with a Young’s modulus $E$ of 200 GPa and a Poisson’s ratio $\nu$ of 0.3. Three uniform quadrilateral meshes were considered: 60x3, 100x5 and 300x15 cells. The  mesh with 100x5 cells is shown in Figure (\ref{fig:scaledCantileverProfile}). The beam is fixed at the left end, by imposing the boundary displacement condition $\overline{\tU} = [0 \; 0]^T$m, and is subjected to a uniform distributed traction at the other end, by imposing the boundary traction condition $\overline{\vv{T}} = [0 \; 1]^T$ MPa. The top and bottom boundaries are traction-free, i.e. $\overline{\vv{T}} = \vv{0}$. Plane strain conditions are assumed.

\begin{figure}
    \centering
	\includegraphics[width=8.0cm, height=6.0cm]
	{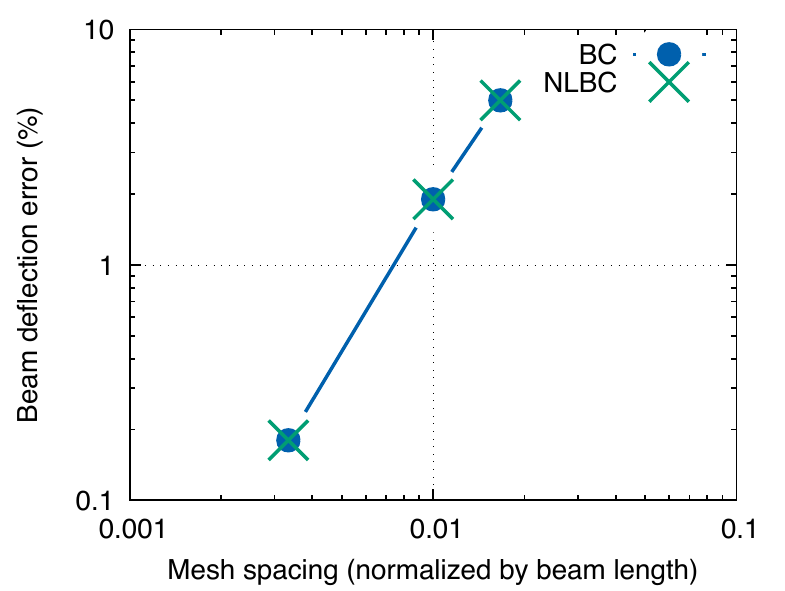}
	\caption{Error in cantilever end-deflection for different mesh refinements. Clearly the approaches match consistently.}
    \label{fig:slenderCantileverMeshRefinement}
\end{figure}

This problem has analytical solution and the deflection on the right-end of the beam is given as \cite{timoshenko1970}: 
\begin{equation}
	\Delta = \frac{PL^3}{3\left(\frac{E}{1-\nu^2}\right)I} = 14.56 \times 10^{-3} \text{m}
\end{equation}
where $P=0.1 \times 10^{6}$ N is the applied load, $L = 2$ m is the length of the beam, and $I=\frac{bh^3}{12} = \frac{0.1^3}{12}\text{ m}^4$ m is the second moment of area of the beam about its bending axis. %
A metric defined as the difference between the predicted displacement and the analytical solution shows that both results from BC and NLBC match consistently (Fig. \ref{fig:slenderCantileverMeshRefinement}), reflecting the analytical proof of equivalence between the formulations inside the boundaries of the linearised elasticity framework.

The NLBC method converged with only one correction step. Finally, just for comparison, the SEG method needs more than 23000 correction steps for mesh 60x3 cells.
%

\subsection{Finite elasticity}
\begin{figure}[!tbp]
  \centering
  \begin{minipage}[b]{0.48\textwidth}
    \includegraphics[width=\textwidth]
    {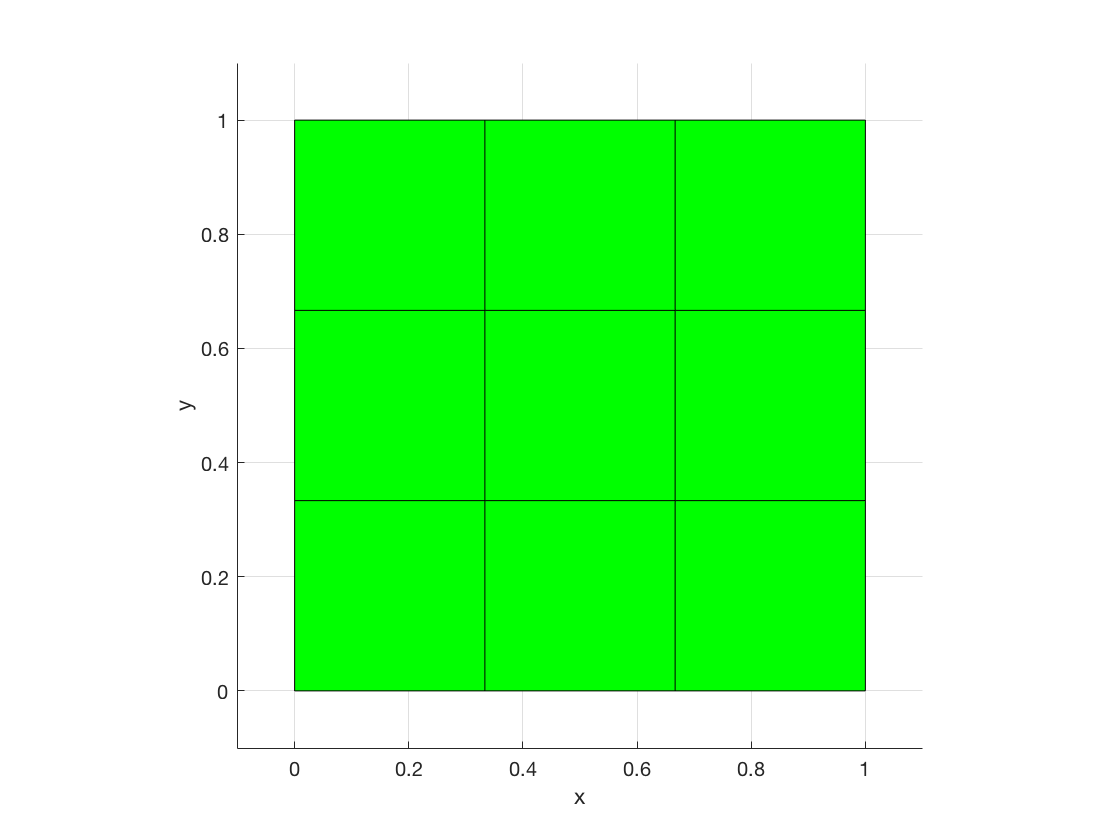}
  \end{minipage}
  \hfill
  \begin{minipage}[b]{0.48\textwidth}
    \includegraphics[width=\textwidth]
    {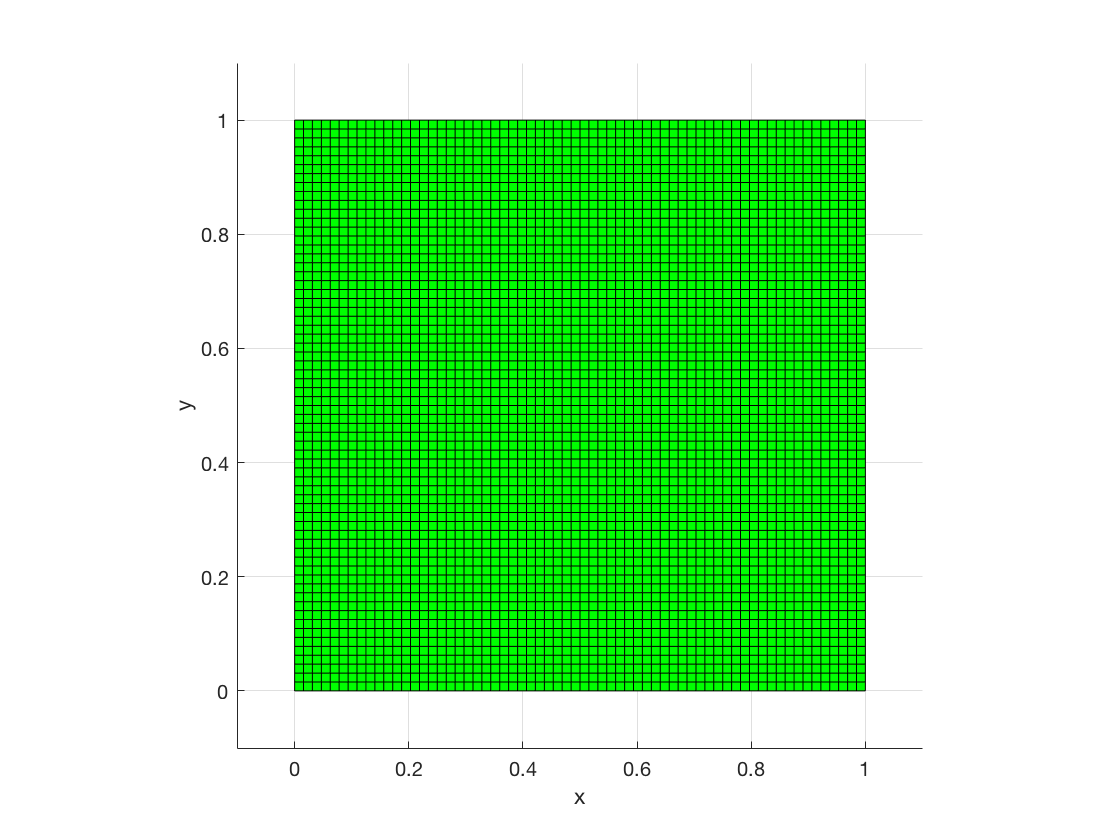}
  \end{minipage}
  \caption{The Coarsest (3$\times$3 cells) and the finest meshes (64$\times$64 cells).}
  \label{fig:unitSquareMeshes}
\end{figure}

Using finite elasticity framework, which allows simulation of accurate “large displacement-large strain” models, it is presented here the comparison of NLBC with the SEG solution procedure. All test cases use the unit square domain and its five uniform discretisation levels. In particular, five Cartesian meshes were considered: 3$\times$3, 8$\times$8, 16$\times$16, 32$\times$32 and 64$\times$64 cells. The coarsest and finest meshes are shown in Figure (\ref{fig:unitSquareMeshes}).

The following metrics were defined to quantify the difference between the predicted displacement and the analytical solution:
\begin{equation}
e_\text{abs}
\left\{
  \begin{array}{lll}
    \mbox{Mean error =} & \frac{1}{n_\text{cells}} 
    	\sum\limits_{i=1}^{n_\text{cells}} r^i \\
    \mbox{Max error =} & \text{max} \{ r^{1}, r^{2}, ..., r^{n_\text{cells}}\} \\
    \mbox{Min error =} & \text{min} \{ r^{1}, r^{2}, ..., r^{n_\text{cells}}\},
  \end{array}
\right.
\end{equation}
where $n_\text{cells}$ is the total number of cells composing the mesh, the sum is over all cells and considering a cell $C_a$, $r^{a} = | \tU_\text{calculated}^{C_a} - \tU_\text{analytic}^{C_a} |$. Every test case was split into two versions: one for displacement-only (Dirichlet) boundary conditions and another for traction-only boundary (Neumann) conditions (except for one boundary, which is set to zero-displacement in order to avoid rigid-body motions). This split scheme isolates patterns which arise due to different boundary condition discretisations employed by NLBC and errors from each one can be investigated individually.

All test cases were created using the accepted standard of verification testing, the Method of Manufactured Solutions (MMS), which allows validation against analytical solution \cite{kamojjala2013}. A MMS test prescribes the deformation map $\deformationMap$, or any other map that allows one to recover it.

The density $\density_0$ was set to 216 $\text{kg/m}^3$; the Young’s Moduli $E$ and Poisson’s ratio $\nu$ were set to 0.02 GPa and 0.3, respectively.

\subsubsection{Uniaxial test cases}
Two homogeneous uniaxial strain MMS were simulated. The deformation gradient $\tF$ is the mapping prescribed for these cases and it is given as:
\begin{equation}
\tF = 
\begin{bmatrix}
    \phi(t) & 0 & 0 \\
    0 & 1 & 0 \\
    0 & 0 & 1
\end{bmatrix}, \quad \text{where} \quad \phi(t) = 1 + (\Lambda - 1)t \quad \text{and} \quad 0 \le t \le 1.
\end{equation}
Note that $\tF$ is homogeneous, i.e. does not depend on a material point $\tRefPosition$. The deformation map is defined as: $\tCurPosition = \deformationMap(\tRefPosition) \equiv \tF \cdot \tRefPosition$ and it is used to set the displacement boundary condition by imposing 
\begin{equation}
	\overline{\tU} = \tCurPosition - \tRefPosition
	\label{eq:dirichletBoundaryCondition}
\end{equation}
at the boundary face centroids. A traction boundary counterpart can be set by noting that a traction $\vv{T}$ acting on the face with unit normal $\N$ is
\begin{equation}
\overline{\vv{T}} = \firstPiolaInGradU(\GradU) \cdot \N = \firstPiolaInGradU(\tF - \I) \cdot \N.
\end{equation}

\subsubsection*{Compression for displacement boundary}
\begin{figure}
  \centering
  \begin{minipage}[b]{0.48\textwidth}
    \includegraphics[width=\textwidth]
    {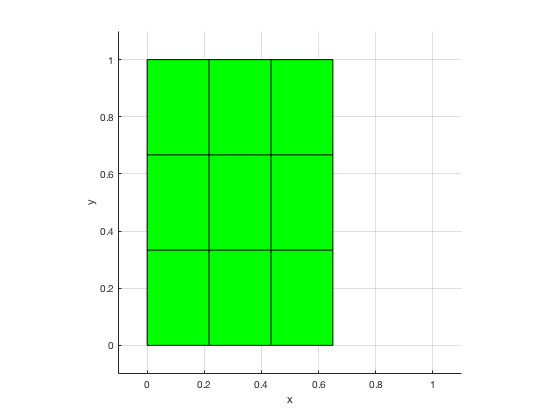}
    \caption{The final deformed domain at compression level $\Lambda = 0.65$.}
    \label{fig:uniaxial2d-065Lambda}
  \end{minipage}
  \hfill
  \begin{minipage}[b]{0.48\textwidth}
    \includegraphics[width=\textwidth]
    {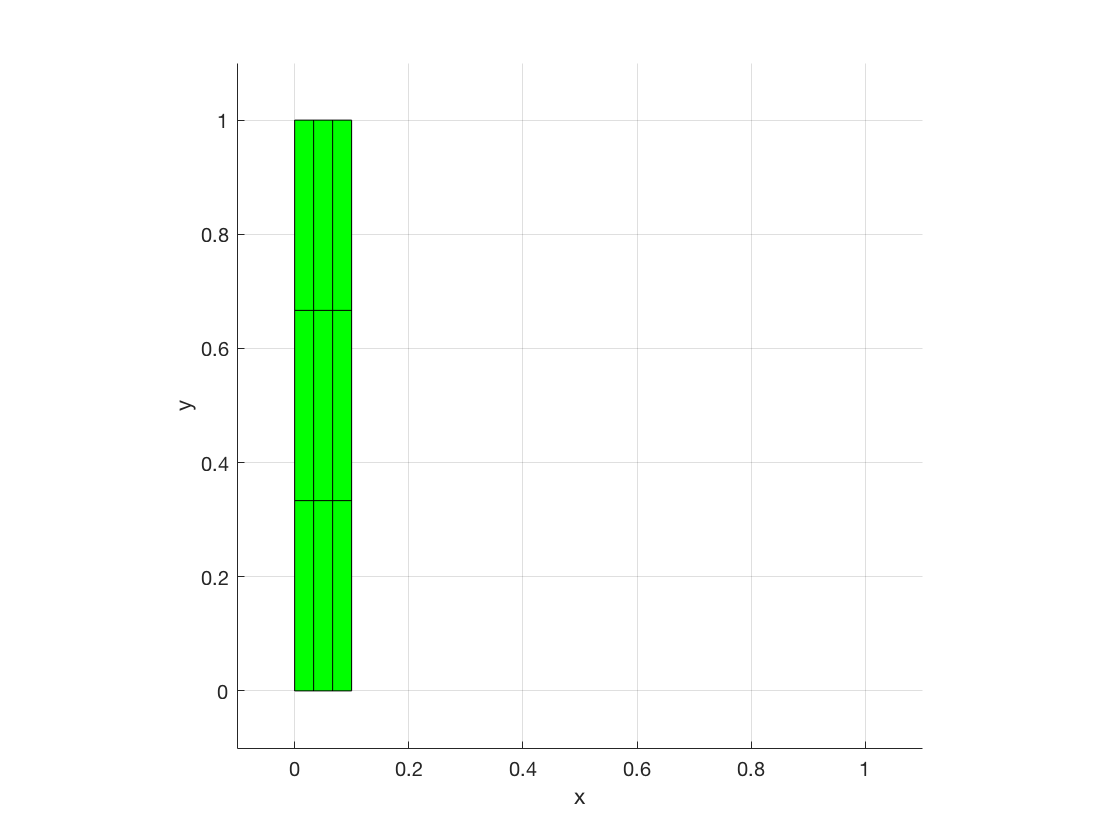}
    \caption{The final deformed domain at compression level $\Lambda = 0.1$.}
    \label{fig:uniaxial2d-01Lambda}
  \end{minipage}
\end{figure}

A variation of the homogeneous uniaxial strain test case described in \cite{kamojjala2013} is presented in this section. However, instead of traction, displacement boundary condition was adopted. Two compression levels were investigated by assigning different values for the compression factor $\Lambda$, in particular, $\Lambda = 0.65$ and $=0.1$ (see Fig. \ref{fig:uniaxial2d-065Lambda} and \ref{fig:uniaxial2d-01Lambda}).

\begin{figure}
  \begin{center}
    \includegraphics
    {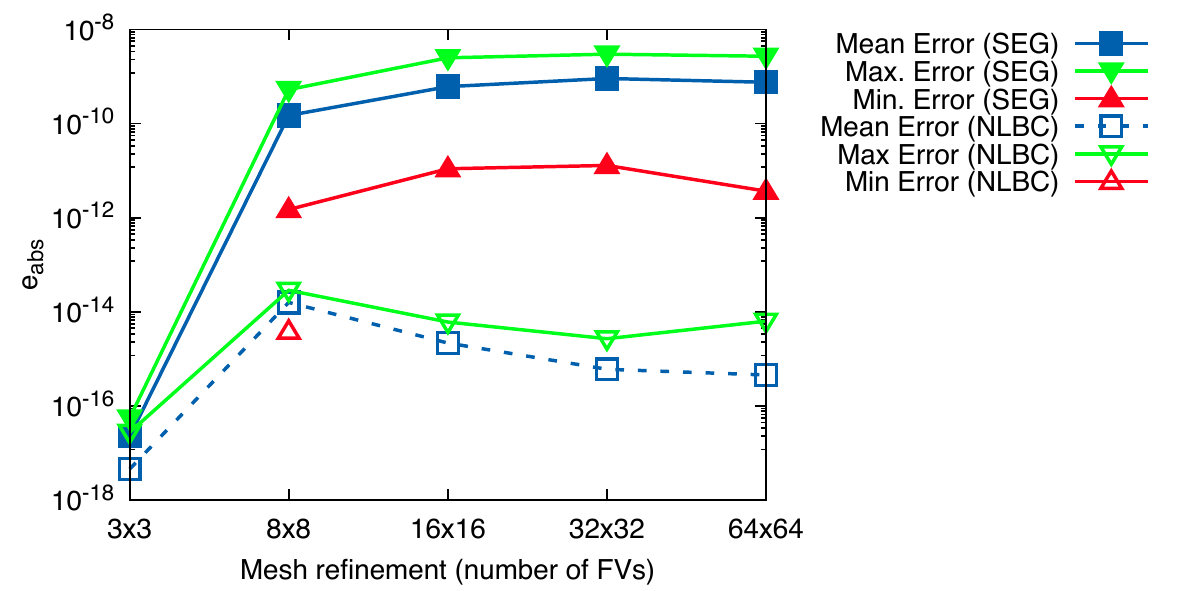}
    \caption{Errors from compression for displacement boundary test case using Neo-Hookean material. The missing data corresponds to when the difference between the solutions is below machine precision.}
    \label{fig:compressionForDisplacementBoundaryLambda0_65}
  \end{center}
\end{figure}

The computed solution with the coarsest mesh was already enough to produce $e_\text{abs} < 10^{-16}$, regardless the method, for $\Lambda = 0.65$ (see Fig. \ref{fig:compressionForDisplacementBoundaryLambda0_65}). The convergence in all scenarios was achieved with only one correction step, i.e. $n_\text{corr} = 1$. When $\Lambda$ is decreased to $0.1$, the SEG method produces $e_\text{abs} < 10^{-9}$. The errors for NLBC also increase when $\Lambda$ get smaller, but they are still relatively small ($e_\text{abs} < 10^{-13}$) and only one correction is needed, considering any mesh.

\subsubsection*{Compression for traction boundary}
Just changing from Dirichlet to Neumann makes the convergence a challenge for both methods, in particular, they are not able to simulate big compression. The summarized results gathered from simulations are:
\begin{itemize}
	\item The SEG method converges only when using the 3$\times$3 cells mesh and $\Lambda \ge 0.8$, but with relatively high errors ($e_\text{abs} > 10^{-2}$).
	\item The NLBC method also converges only for $\Lambda \ge 0.8$ and provided that meshes are more refined than or equal to the mesh 16$\times$16. For these scenarios, $e_{\text{abs}} < 10^{-7}$.
\end{itemize}

\subsubsection*{Tension for displacement boundary}
\begin{figure}
  \begin{center}
      \includegraphics[width=0.48\linewidth]{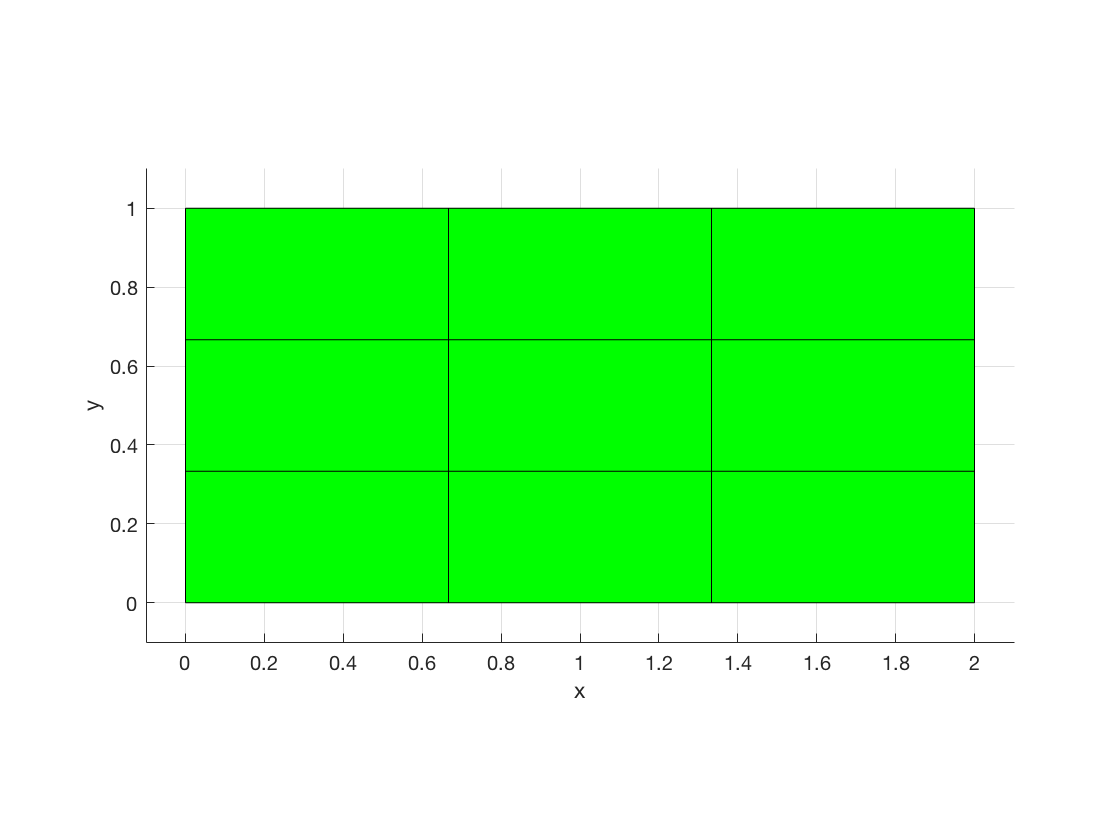}
      \caption{The final deformed domain for tensile strain case and for displacement boundary condition.}
      \label{fig:uniaxial2d_20Lambda}
  \end{center}
\end{figure}

The cases above were repeated, but with $\Lambda > 1$ in order to simulate tension, in particular, $\Lambda = 2$ was adopted. The Figure (\ref{fig:uniaxial2d_20Lambda}) shows the final deformed domain for the coarsest mesh.

\begin{figure}
  \begin{center}
      \includegraphics
      {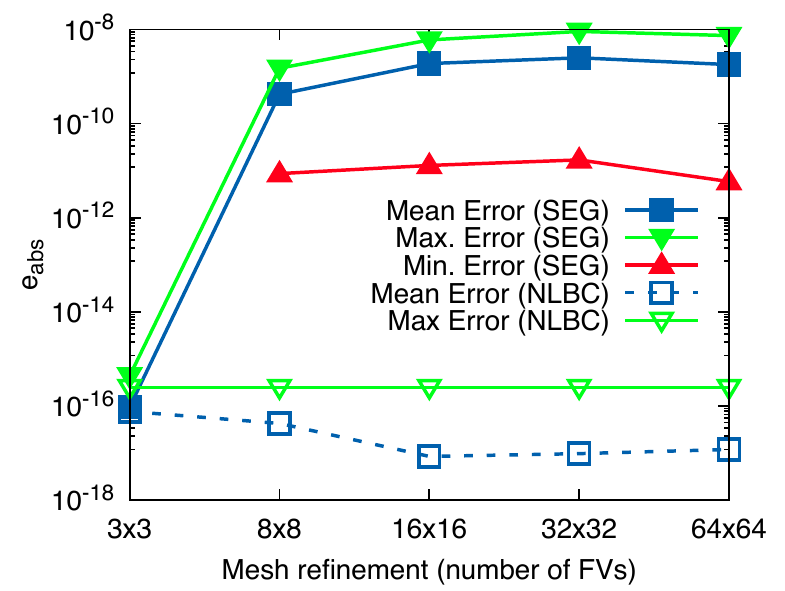}
      \caption{Errors from tension for displacement boundary test case. The missing data corresponds to when the difference between the solutions is below machine precision.}
      \label{fig:SegBLNCUniaxialNH2_0LmdDisp1ts200corrnFVMMeshRef}
  \end{center}
\end{figure}

Interestingly, something changes when tension is simulated. Both methods converge for all meshes with errors $e_\text{abs} < 10^{-8}$ (see Fig. \ref{fig:SegBLNCUniaxialNH2_0LmdDisp1ts200corrnFVMMeshRef}). Note that NLBC produces significantly smaller errors. Both methods converge with only one correction step.

\subsubsection*{Tension for traction boundary}
\begin{figure}
  \begin{center}
      \includegraphics
      {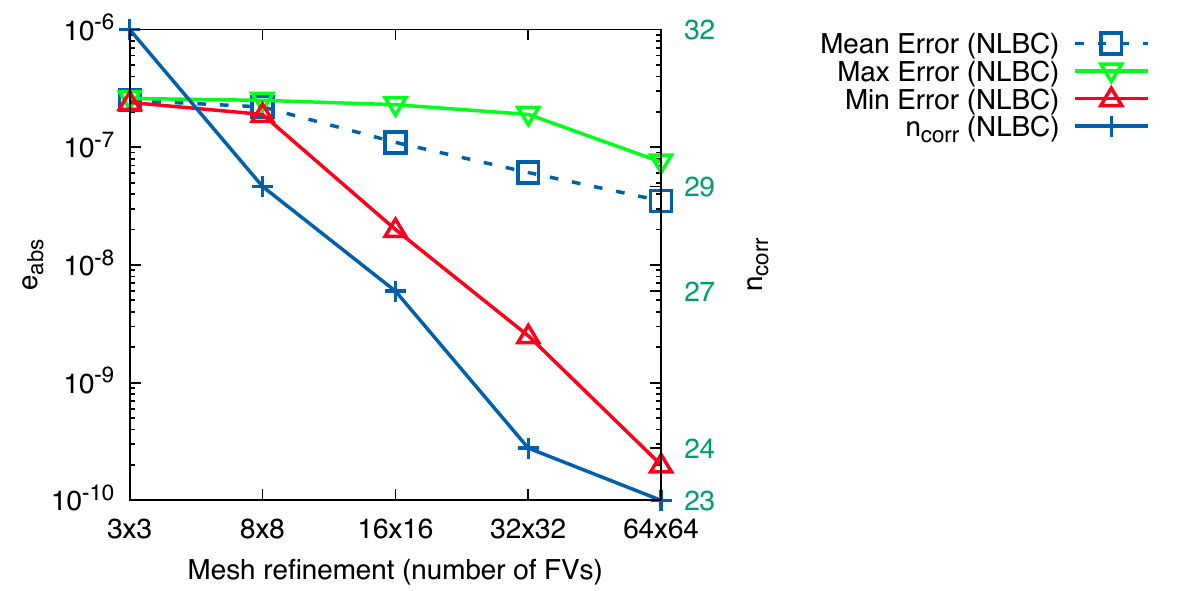}
      \caption{Error in tension for traction boundary test case as mesh is refined. The $n_\text{corr}$ as mesh is refined is also shown. Results are only for NLBC, since SEG method could not handle this case.}
      \label{fig:uniaxial2dTraction2_0Lmd_nFVMMeshRef}
  \end{center}
\end{figure}

Once again, when traction is introduced, SEG does have convergence problems. In fact, it does not converge for $\Lambda$ much greater than one. And even when $\Lambda$ is close to one, e.g. 1.2, the errors are relatively high (either with nFVM or S4F). Regarding NLBC's results, they show good agreement with analytical solution. The method converges for all meshes, for any $\Lambda \in (1, 2]$ and the errors are relatively small ($e_\text{abs} < 10^{-6}$), but much higher than the corresponding test which uses displacement boundaries (see Fig. \ref{fig:uniaxial2dTraction2_0Lmd_nFVMMeshRef} and compare with Fig. \ref{fig:SegBLNCUniaxialNH2_0LmdDisp1ts200corrnFVMMeshRef}).

\subsubsection{Shear test cases}
\begin{figure}
   \begin{center}
      \includegraphics[width=0.48\linewidth]{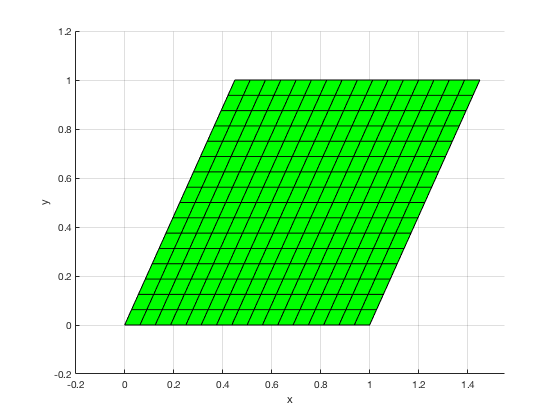}
      \caption{Deformed profile for mesh 16$\times$16 in shear test case.}
      \label{fig:shear2dNLBCNH0_45sfTraction16x16}
  \end{center}
\end{figure}

This test case consists of a simple shear \cite{bonet2008}. The deformation gradient for this manufactured solution is similar to that of the uniaxial test case and is given by (being the shear factor $\omega = 0.45$ chosen arbitrarily):
\begin{equation}
	\F = 
	\begin{bmatrix}
	    1 & \phi(t) & 0 \\
	    0 & 1 & 0 \\
	    0 & 0 & 1
	\end{bmatrix}, \quad \text{where} \quad \phi(t) = \omega t \quad \text{and} \quad 0 \le t \le 1.
	\label{eq:shearMatrix}
\end{equation}
 The Figure (\ref{fig:shear2dNLBCNH0_45sfTraction16x16}) shows the deformed profile for mesh 16$\times$16. The boundary condition is imposed in the same manner as it was done in uniaxial test cases.

\subsubsection*{Shear for displacement boundary}
\begin{figure}
    \begin{center}
        \includegraphics
        {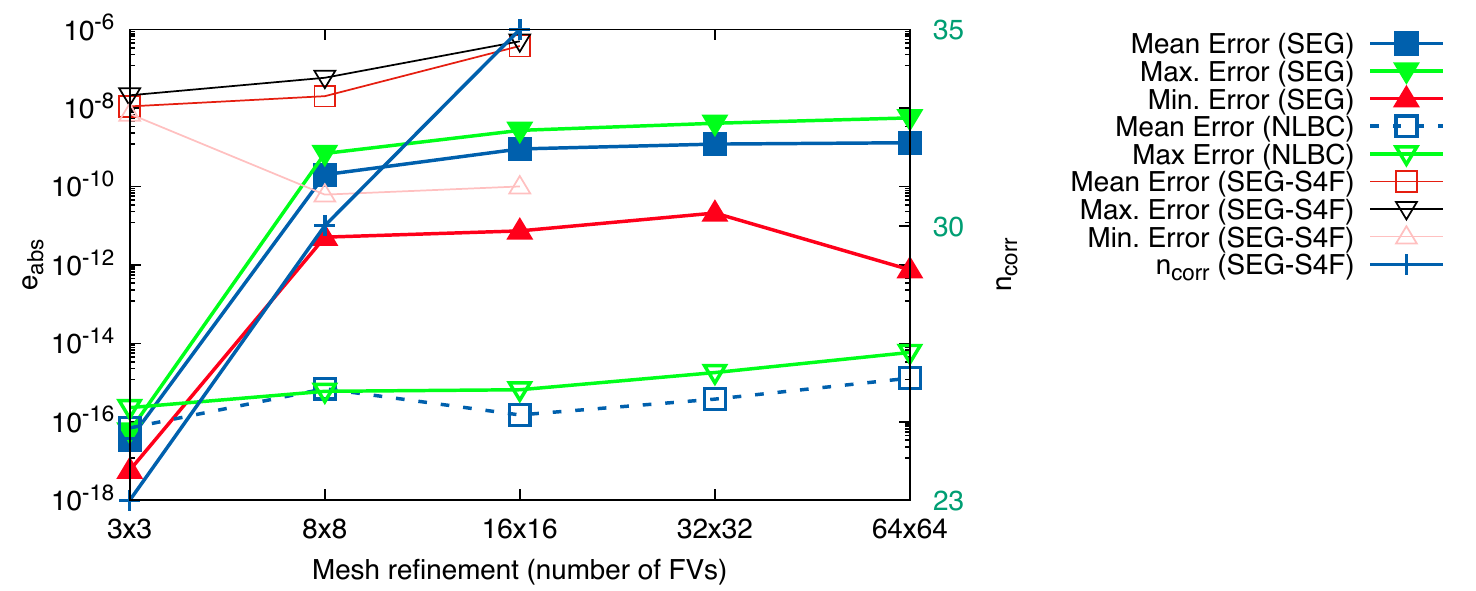}
        \caption{Error in shear for displacement boundary test case as mesh is refined. The number of correction $n_\text{corr}$ as mesh is refined for S4F's SEG is also shown. The SEG and NLBC implementations in nFVM needed only one correction to achieve convergence.}
        \label{fig:shear2dNH0_45sfDisp}
    \end{center}
\end{figure}

The results from simulations were qualitatively similar to that of the uniaxial compression, or tension, for displacement boundary (compare Fig. \ref{fig:compressionForDisplacementBoundaryLambda0_65} and \ref{fig:SegBLNCUniaxialNH2_0LmdDisp1ts200corrnFVMMeshRef} with Fig. \ref{fig:shear2dNH0_45sfDisp}). The nFVM's SEG and NLBC needed only one correction for all meshes. The S4F's SEG was also tested and the output shows that as mesh gets refined, it needs more corrections to achieve convergence ($n_\text{corr}=23, 30$ and $35$ for meshes 3$\times$3, 8$\times$8, 16$\times$16 respectively). Besides, it did not converge for meshes finer than 16$\times$16.

\subsubsection*{Shear for traction boundary}
\begin{figure}
    \centering
    \includegraphics
    {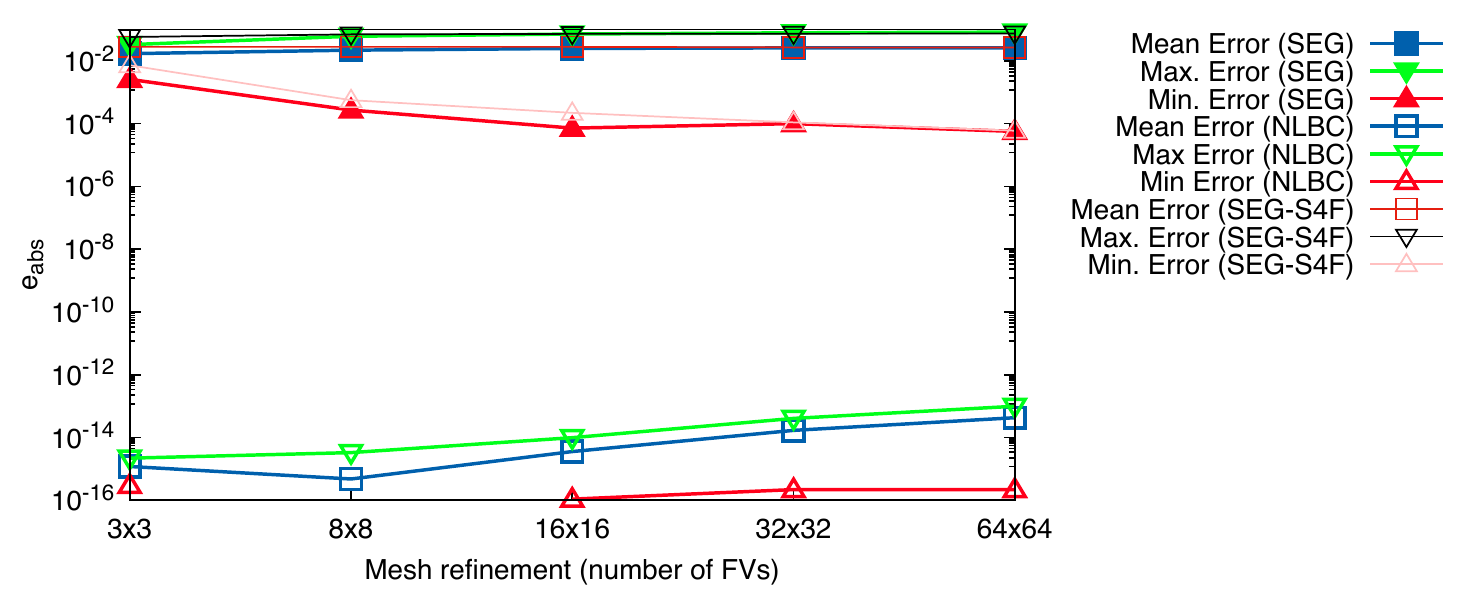}
    \caption{Error in shear for traction boundary test case as mesh is refined.}
    \label{fig:shear2dNH0_45sfTraction}
\end{figure}

\begin{figure}
    \centering
    \includegraphics
    {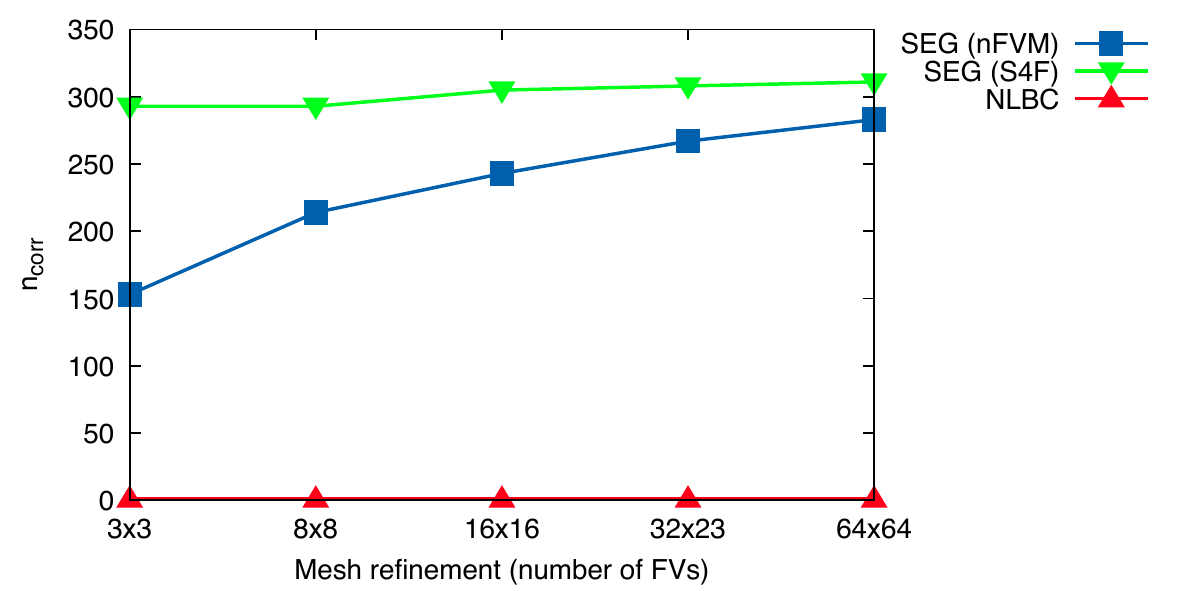}
    \caption{Number of corrections in shear for traction boundary test case as mesh is refined. The $n_\text{corr}$ for NLBC is 1.}
    \label{fig:shear2dNH0_45sfTractionNcorr}
\end{figure}

Once more, when displacement boundaries are replaced by traction boundaries, the SEG method has convergence problems (both in nFVM and in S4F). In fact, convergence is achieved, however with relatively high errors (see Fig. \ref{fig:shear2dNH0_45sfTraction}). The SEG approach needs more than 100 correction steps to converge and for the finer the mesh, more correction steps are necessary for convergence (see Fig. \ref{fig:shear2dNH0_45sfTractionNcorr}).

The results from the NLBC method were in good agreement with analytical solutions and only one correction was needed in order to achieve convergence (see Fig. \ref{fig:shear2dNH0_45sfTractionNcorr}) using any mesh. 

\begin{figure}
    \centering
    \includegraphics{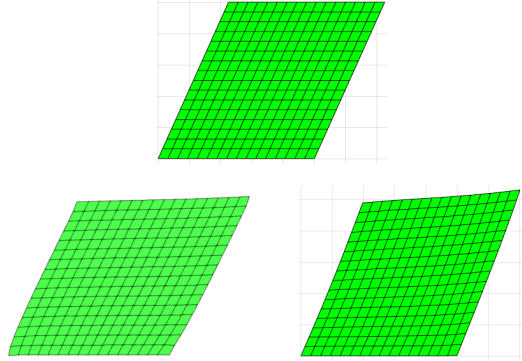}
    \caption{Deformed profiles for SEG using S4F (lower-left corner) and nFVM (lower-right corner). On top the result from NLBC. Traction boundary was used.}
    \label{fig:SegShear2dNH0_45sfTraction1ts250corr16x16S4F_and_nFVM}
\end{figure}

The final deformation domain (Fig. \ref{fig:SegShear2dNH0_45sfTraction1ts250corr16x16S4F_and_nFVM}) for SEG is clearly ``warped'' (and refining the mesh does not reduce this spurious artifact).

\section{Discussion \& conclusions}
\label{sec:conclusion}
It has been presented a novel nonlinear block-coupled FV methodology which generalises the work of Cardiff et al. \cite{CARDIFF2016100}. The developed methodology has been investigated by means of numerical simulations, i.e. one test case for infinitesimal elasticity and four for finite elasticity. The accuracy of the methodology has been shown through detailed comparison with analytical solutions and numerical benchmarks. For all the test cases analysed, NLBC has shown to be an efficient and accurate alternative to SEG method for the analysis of 2-D problems in finite elasticity.

The novel methodology does not assume small strain or small displacement during the discretisation process and only requires the presence of a right-minor symmetric elasticity tensor. Thus, it defines a class of “officially” supported solid models. In fact, it can be demonstrated that a large set of important solid models (those that are hyperelastic, frame-indifferent, homogeneous and isotropic) belong to this class. As a matter of fact, frame-indifference should be required independently of the FVM methodology, since it is required in finite elasticity, otherwise different observers could collect different results \cite{bonet2008}. This symmetric-related elasticity restriction should be subjected to investigation in order to establish if it can be removed or at least weakened.

\ifodd 2{
- Much superior than BC (not restricted to linearised elasticity);
- Much superior than SEG (convergence, convergence rate and errors);
- Traction discretisation problems are case-dependent.

[From thesis presentation]
- The creation of new FVM methodology that closely resembles FEM
- It generalizes the BC methodology;
- The NLBC method has shown to be superior when comparing with BC, since the latter is restricted to linearised elasticity;
- The rate of convergence of NLBC is really impressive, i.e. when compared to that of SEG’s; (I don't have the timings, thus forget it!)
- Closer to the long awaited unification for FSI;
- In all test cases NLBC approach was much superior than SEG approach, i.e. in regard to convergence rate and divergence;
- Test cases without traction boundary condition, when converging for all time steps, NLBC does that with relatively small errors;
- The NLBC’s convergence problems from traction discretisation is case-dependent;

[From thesis conclusion]
- The tests clearly showed that the adopted traction boundary condition discretisation needs to be improved;
- The “borrowed” (from BC) traction boundary discretisation for NLBC seems to cause divergence.

[From BC,Cardiff - 2016]
- The new method provides an efficient and accurate alternative to segregated FV methods and standard low order FE methods for the analysis of complex 3-D problems in linear elasticity, and has considerable potential to be extended to multi-physics simulations.
}\else{}\fi
As regards mesh support, NLBC assumes that finite volumes are rectangular cuboids. However, by judging how other FV methodologies handle convex polyhedral, the modification to add support to it should be relatively straight-forward.

In conclusion, it has been presented the first attempt to generalise the BC solution methodology to finite elasticity, for which a Newton-Raphson method was employed similar as in finite element analysis.

In conclusion, it has been presented the first attempt to generalise the BC solution methodology to finite elasticity, which closely resembles the Finite Element Methodology in the sense that all displacement components are solved at the same time in a big linear system generated by applying the Newton-Raphson procedure on an out-of-balance force function.

\section{Acknowledgments}
This work was supported by the program Ciência sem Fronteiras (Grant 233309/2014-4, CNPq, Brazil) and the ``Excellence Initiative'' of the German Federal and State Governments and the Graduate School of Computational Engineering at Technische Universität Darmstadt.

\appendix
\section{Elastic body}
\label{elasticBody}
An elastic body can be defined through the following \textbf{elastic body axiom} \cite{gonzalez2008}: a continuum body with reference configuration $\body$ is elastic if $\exists \; \tCauchyInF : \euclideanVectorSpace^2 \times \body \to \euclideanVectorSpace^2$ such that
\begin{equation}
	\begin{split}
		&\tCauchy_m(\vv{X},t) = \tCauchyInF(\tF(\vv{X},t), \vv{X}), \quad \forall \vv{X} \in \body, t \geq 0 \quad \text{and} \\
		&\tCauchyInF(\tF,\vv{X})^T = \tCauchyInF(\tF,\vv{X}), \quad \forall \vv{X} \in \body, \tF \in \euclideanVectorSpace^2, \deter{\tF} > 0.
	\end{split}
	\label{eq:elasticBodyDef}
\end{equation}
Since this work considers only homogeneous bodies, the \textbf{stress response function} $\tCauchyInF$ is considered independent of $\vv{X}$, thus $\tCauchy_m(\vv{X},t) = \tCauchyInF(\tF(\vv{X},t))$. Because of this axiom, there are two functions 
$\firstPiolaInF : \euclideanVectorSpace^2 \to \euclideanVectorSpace^2$ and 
$\secondPiolaInF : \euclideanVectorSpace^2 \to \euclideanVectorSpace^2$ such that
\begin{equation}
	\tFirstPiola(\vv{X},t) = \firstPiolaInF(\tF(\vv{X},t))
	\quad \text{and} \quad
	\tSecondPiola(\vv{X},t) = \secondPiolaInF(\tF(\vv{X},t)),
	\label{eq:fstAndSecpiolaHatImplication}
\end{equation}
in particular, they must satisfy the relations
\begin{equation}
		\firstPiolaInF(\tF) = (\deter{\tF}) \tCauchyInF(\tF) \cdot \tF^{-T}
		\quad \text{and} \quad
		\secondPiolaInF(\tF) = \tF^{-1} \cdot \firstPiolaInF(\tF).
	\label{eq:fstAndSecpiolaHatImplicationRelations}
\end{equation}

The \textbf{axiom of material frame-indifference} implies that: $\exists \; \secondPiolaInC : \euclideanVectorSpace^2 \to \euclideanVectorSpace^2$ such that
\begin{equation}
		\firstPiolaInF(\tF) = \tF \cdot \secondPiolaInC(\tC)
		\quad \text{and} \quad
		\secondPiolaInF(\tF) = \secondPiolaInC(\tC)
	\label{eq:materialFrameIndifferenceAxiom}
\end{equation}
where $\tC = \tF^T \cdot \tF$ is the right Cauchy-Green deformation tensor.

Let $\tA \in \euclideanVectorSpace^2$, then
\begin{equation}
	\FstDev[\tC]{\tF} : \tA = \tA^T \cdot \tF + \tF^T \cdot \tA,
\end{equation}
thus (using the chain rule and $\partial \tF / \partial \GradU = \tFthI$, i.e. the fourth-order identity tensor
\footnote{The definition is 
$\tFthI
	\equiv \delta_{ac} \delta_{bd}
	\eBase_{a} \otimes \eBase_{b} \otimes \eBase_{c} \otimes \eBase_{d} 
$ which implies that $\tFthI : \vv{A} = \vv{A}$.})
\begin{equation}
	\begin{split}
		\FstDev[\tC]{\GradU} : \tA 
		&= \FstDev[\tC]{\tF} : \bigg{(} \FstDev[\tF]{\GradU} : \tA \bigg{)} = \FstDev[\tC]{\tF} : \bigg{(} \tFthI : \tA \bigg{)} = \FstDev[\tC]{\tF} : \tA \\
		&= \tA^T \cdot \tF + \tF^T \cdot \tA \equiv 2 \cdot \sym{(\tF^T \cdot \tA)},
	\end{split}
	\label{eq:dCdGradU}
\end{equation}
where $\sym{(\cdot)}$ denotes the symmetric component of a tensor
\footnote{The last equation shows that 
$\sym{(\tensor{B})} \equiv \frac{1}{2}(\tensor{B} + \tensor{B}^T)$.}. The field $\partial \tC / \partial \GradU$ is used next.

The Green-Lagrange strain tensor $\E = \frac{1}{2} ( \tC - \I )$ and the new function $\secondPiolaInE(\E) = \secondPiolaInC(\tC(\E))$ can be used to find the \textbf{elasticity tensor} 
$\tElasticityTensor = \partial \secondPiolaInE/\partial \E$ in terms of 
$\partial \secondPiolaInC/\partial \tC$ as
\begin{equation}
	\begin{split}
		\FstDev[ 
			\secondPiolaInE
		]{\E} : \tA 
		&= 
		\FstDev[ 
			\secondPiolaInC
		]{\tC} :
		\bigg{(} \FstDev[\tC]{\E} : \tA \bigg{)} 
		 \quad \quad \text{(using again the chain rule)} \\
		&= 
		\FstDev[ 
			\secondPiolaInC
		]{\tC} :
		\bigg{(} 2 \tFthI : \tA \bigg{)} \\
		&= 
		2
		\FstDev[ 
			\secondPiolaInC
		]{\tC} : \tA.
	\end{split}
\end{equation}
The arbitrariness of $\tA$ implies that
\begin{equation}
	\tElasticityTensor 
	= 
	\FstDev[ 
		\secondPiolaInE
	]{\E}
	=
	2 \FstDev[ 
		\secondPiolaInC
	]{\tC}.
	\label{eq:elasticityTensor}
\end{equation}

Using Equation (\ref{eq:materialFrameIndifferenceAxiom}) and the definition of the two new functions $\firstPiolaInGradU(\GradU) = \firstPiolaInF(\tF(\GradU)) = \firstPiolaInF(\I + \GradU)$ and $\tSecondPiolaInGradU(\GradU) = \secondPiolaInC(\tC(\GradU))$, the derivative of the stress response function $\firstPiolaInGradU$ is given as
\begin{equation}
	\begin{split}
		\FstDev[\firstPiolaInGradU]{\GradU} : \tA
		&= 
		\FstDev[ 
			(
				\tF \cdot \tSecondPiolaInGradU
			)
		]{\GradU} : \tA \quad \quad \text{(using Equation (\ref{eq:materialFrameIndifferenceAxiom}))} \\
		&=
		\bigg{(}
			\FstDev[ 
				\tF
			]{\GradU} : \tA
		\bigg{)} \cdot \tSecondPiolaInGradU
		+
		\tF \cdot
		\bigg{(}
			\FstDev[ 
				\tSecondPiolaInGradU
			]{\GradU} : \tA
		\bigg{)} \\
		&=
		\tA \cdot \tSecondPiolaInGradU
		+
		\tF \cdot
		\bigg{(}
			\FstDev[ 
				\tSecondPiolaInGradU
			]{\GradU} : \tA
		\bigg{)} \quad \quad \text{(using } \partial \tF / \partial \GradU = \tFthI)  \\
		&=
		\tA \cdot \tSecondPiolaInGradU
		+
		\tF \cdot
		\bigg{[}
			\FstDev[ 
					\secondPiolaInC
				]{\tC}
			:
			\bigg{(}
				\FstDev[ 
					\tC
				]{\GradU} : \tA
			\bigg{)}
		\bigg{]} 
		\quad \quad \text{(using the chain rule)} \\
		&=
		\tA \cdot \tSecondPiolaInGradU
		+
		\tF \cdot
		\bigg{[}
			\FstDev[ 
					\secondPiolaInC
				]{\tC}
			:
			\bigg{(}
				2 \cdot \sym{(\tF^T \cdot \tA)}
			\bigg{)}
		\bigg{]} \quad \quad \text{(using Eq. (\ref{eq:dCdGradU}))} \\
		&=
		\tA \cdot \tSecondPiolaInGradU
		+
		\tF \cdot
		\bigg{[}
			\tElasticityTensor
			:
			\bigg{(}
				\sym{(\tF^T \cdot \tA)}
			\bigg{)}
		\bigg{]} \quad \quad \text{(using Eq. (\ref{eq:elasticityTensor}))}.
	\end{split}
	\label{eq:dPdGradU}
\end{equation}
The last equation needs to be extended by taking the $\tElasticityTensor$'s right-minor symmetry into consideration (2.51) as
\begin{equation}
	\begin{split}
		\FstDev[\firstPiolaInGradU]{\GradU} : \tA
		&=
			\tA \cdot \tSecondPiolaInGradU
			+
			\tF \cdot
			\bigg{[}
				\tElasticityTensor
				:
				\bigg{(}
					\sym{(\tF^T \cdot \tA)}
				\bigg{)}
			\bigg{]} \\
		&=
			\tA \cdot \tSecondPiolaInGradU
			+
			\tF \cdot
			\bigg{[}
				\tElasticityTensor : \bigg{(} \tF^T \cdot \tA \bigg{)}
			\bigg{]} 
			\quad \quad \text{(using $\tElasticityTensor$'s symmetric property)}\\
		&=
			\tA \cdot \tSecondPiolaInGradU
			+
			\tF \cdot
			\bigg{[}
				\cElasticityTensor_{\alpha \beta \gamma \delta}
				\cF_{a\gamma}
				\cA_{a\delta}
				\eBase_{\alpha} \otimes \eBase_{\beta}
			\bigg{]} \\
		&=
			\tA \cdot \tSecondPiolaInGradU
			+
			\bigg{(}
				\tF \cdot \tElasticityTensor \underset{(3)}{\cdot} \tF^T
			\bigg{)} : \tA \\
		&=
			\tA \cdot \tSecondPiolaInGradU
			+
			\tTransformedElasticityTensor : \tA,
			\quad\quad \forall \tA \in \euclideanVectorSpace^2,
	\end{split}
	\label{eq:firstPiolaInGradUInnerA}
\end{equation}
where $\tTransformedElasticityTensor$ is the transformed elasticity tensor defined in the paragraph preceding Equation (2.53). Note that the right-minor symmetry restriction, which could not be overcome, creates a class of supported materials.

\section{Neo-Hookean model}
\label{neoHookean}
This is a compressible isotropic hyperelastic material model and its strain-energy function is defined as \cite{bonet2008,doghri2013}
\begin{equation}
\strainEnergyInrCG(\tC) = \strainEnergyInInvar(
		\cFthTsr{I}_{\tC}
	) = \frac{\mu}{2}(\firstTensorInvariantFunction{\tC} - 3) - \mu \ln J + \frac{\lambda}{2} (\ln J)^2,
	\label{eq:strainEnergyNH}
\end{equation}
where $\thirdTensorInvariantFunction{\tC} = \deter{\tC} = J^2$. The Lamé (material) coefficients $\lambda$ and $\mu$ relating to the Young’s modulus $E$ and Poisson's ratio, $\nu$, are given respectively as: \( \displaystyle \mu = \frac{E}{2(1+\nu)} \);
\(\displaystyle \frac{\nu E}{(1+\nu)(1-\nu)}\) for plane stress; and \(\displaystyle \frac{\nu E}{(1+\nu)(1-2\nu)}\) for plane strain and 3-D.
The second Piola-Kirchhoff stress tensor is obtained from Equation (\ref{eq:strainEnergyNH}) as
\begin{equation}
	\tSecondPiola
	  	= \secondPiolaInC(\tC)
	  	= 2 \FstDev[\strainEnergyInrCG(\tC)]{\tC}
	  	= \mu(\I - \tCinv) + \lambda (\ln J) \tCinv.
	\label{eq:SndPiolaForNH}
\end{equation}
\ifodd 2 {
Using
\begin{equation}
	\tFirstPiola
	=  (\det \tF) \tCauchy_m \cdot \tF^T
\end{equation}
and
\begin{equation}
	\tSecondPiola
	= \tF^{-1} \cdot \tFirstPiola
\end{equation}
it is straightforward to prove that
\begin{equation}
	\tCauchy_m = \mu J^{-1}(\tF \cdot \tF^{T} - \I) + \lambda (\ln J) J^{-1} \I
\end{equation}
}\else{}\fi
The elasticity tensor can be obtained by differentiation of Equation (\ref{eq:SndPiolaForNH}) with respect to the components of $\E$ to give, after some algebra using $\partial \thirdTensorInvariantFunction{\tC}/\partial \tC = J^2 \tCinv$, $\tElasticityTensor$ as
\begin{equation}\label{eq:elsTensorNH}
	\tElasticityTensor = 
			\FstDev[\secondPiolaInE]{\E}
		= 2 \FstDev[\secondPiolaInC]{\tC}
		= \lambda \tCinv \otimes \tCinv + 2(\mu - \lambda \ln J) \tJ,
\end{equation}
where the fourth-order tensor $\tJ$ is defined as 
\begin{equation}\label{eq:tJ}
  \tJ = - \FstDev[\tCinv]{\tC}
  \iff
  \cJ_{IJKL} = \frac{1}{2}
  \Big{[} 
    (C^{-1})_{IK} (C^{-1})_{JL} + (C^{-1})_{IL} (C^{-1})_{JK}
  \Big{]}.
\end{equation}
It is straightforward to show that $\tJ$, $\tCinv \otimes \tCinv$ (using $(\cC^{-1})_{KL} = (\cC^{-1})_{LK}$) and therefore the elasticity tensor $\tElasticityTensor$ above has right-minor symmetry, i.e.
\begin{equation}
	\cElasticityTensor_{IJKL} = \cElasticityTensor_{IJLK}.
\end{equation}

The full expression for $\tT^d$ is obtained by substituting Equation (\ref{eq:elsTensorNH}) into Equation (\ref{eq:tdDefinition}) resulting in
\begin{equation}
  \begin{split}
    \tT^d &= \lambda (\tF \cdot \tC^{-1} \cdot \N) \otimes (\vv{f}^d \cdot \tC^{-1})\\ 
          &\quad + (\mu - \lambda \ln J)
         \Big{[} 
           ( \tF \cdot \tC^{-1} \cdot \vv{f}^d ) \otimes (\N \cdot \tC^{-1})
          + (\N \cdot \tC^{-1} \cdot \vv{f}^d ) 		(\tF \cdot \tC^{-1} )
         \Big{]}\\
          &= \lambda (\tensor{A} \cdot \N) \otimes (\vv{f}^d \cdot \tC^{-1})
         + (\mu - \lambda \ln J)
         \Big{[} 
           ( \tensor{A} \cdot \vv{f}^d ) \otimes \vv{b}
          + (\vv{b} \cdot \vv{f}^d ) \tensor{A}
         \Big{]} \\
         &= \lambda (\tensor{A} \cdot \N) \otimes (\tC^{-1} \cdot \vv{f}^d)
         + (\mu - \lambda \ln J)
         \Big{[} 
           ( \tensor{A} \cdot \vv{f}^d ) \otimes \vv{b}
          + (\vv{b} \cdot \vv{f}^d ) \tensor{A}
         \Big{]} \quad\quad \text{($\tC^{-1}$ is symmetric)},
  \end{split}
  \label{eq:neoHookeansTd}
\end{equation}
where $\tensor{A} = \tF \cdot \tC^{-1}$ and $\vv{b} = \N \cdot \tC^{-1} = \tC^{-1} \cdot \N$.

\section{Mathematical framework for incremental description}
\label{incrDrescription}
\begin{figure}
	\begin{center}
	    \includegraphics
	    {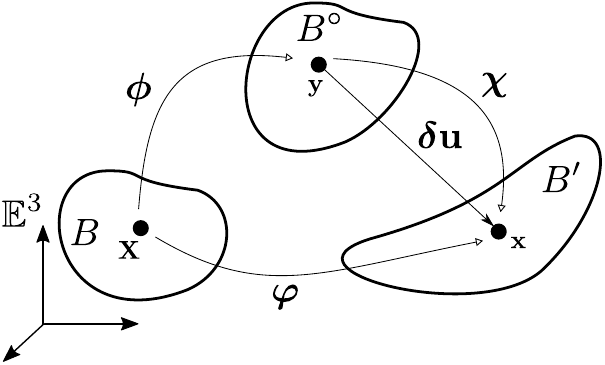}
	    \caption{A deformation is illustrated by considering the reference configuration $\body$, the old configuration $\body^{\circ}$ and the current deformed configuration $\body'$. The black dots represent one and the same material particle.}
	    \label{fig:NLBCdeformationMaps}
    \end{center}
\end{figure}

To describe the incremental approach, let the following maps be defined:
\begin{equation}
	\begin{split}
		\tDeformationMap 	: \vv{X} \in \body &\to \body' \ni \vv{x}, \\
		\tInteDeformationMap : \vv{X} \in \body &\to \body^{\circ} \ni \vv{y}
		\quad \text{and} \\
		\tIncrDeformationMap : \vv{y} \in \body^{\circ} &\to \body' \ni \vv{x},
	\end{split}
\end{equation}
where $\body^{\circ}$ can be thought as an intermediate (also labeled as old) body state between the reference body state $\body$ and the current body state $\body'$ (see Fig. \ref{fig:NLBCdeformationMaps}). Then, by using the composition $\tIncrDeformationMap \circ \tInteDeformationMap$ , it is derived the relation between the deformation gradients associated with the mappings as
\begin{equation}
	\begin{split}
		\vv{x}
		= \tDeformationMap(\vv{X}) 
		= \tIncrDeformationMap (\tInteDeformationMap(\vv{X}))
	 	\implies
	 		\underbrace{\FstDev[\tDeformationMap]{\vv{X}}}_{\tF}
	 	= 
	 		\underbrace{\FstDev[\tIncrDeformationMap]{\vv{y}}}_{\tDeltaF}
	 		\cdot
	 		\underbrace{\FstDev[\tInteDeformationMap]{\vv{X}}}_{\tF^{\circ}},
	 \end{split}
\end{equation}
thus obtaining the relation between the deformation gradients as
\begin{equation}
	\tF = \tDeltaF \cdot \tF^{\circ}.
	\label{eq:relationBetweenDeformationGrads}
\end{equation}
The symbol $\tDeltaF$ is the well known \cite{ogden1997} \textbf{incremental} (or \textbf{relative}) \textbf{deformation gradient}, and its relation with the so-called \textbf{incremental} (or \textbf{relative}) \textbf{displacement gradient} $\tGradDU$ is found using the \textbf{incremental displacement} field $\tDeltaU : \body^\circ \to \body'$ (Fig. \ref{fig:NLBCdeformationMaps}) as:
\begin{equation}
	\tDeltaU(\vv{y}) = \tIncrDeformationMap(\vv{y}) - \vv{y}
	\implies 
	\underbrace{\FstDev[\tDeltaU]{\vv{y}}}_{\tGradDU}
		= \FstDev[\tIncrDeformationMap]{\vv{y}} - \FstDev[\vv{y}]{\vv{y}}
		= \tDeltaF - \I,
\end{equation}
therefore
\begin{equation}
	 \I + \tGradDU = \tDeltaF.
	\label{eq:incrDisplacementGradient}
\end{equation}

The intermediate (or old) displacement field $\tU^{\circ}(\vv{X}) = \tInteDeformationMap(\vv{X}) - \vv{X}$ gives rise to the intermediate (or old) displacement gradient
\begin{equation}
	\begin{split}
		\nabla \tU^{\circ} = \nabla \tInteDeformationMap - \I
		\implies \nabla \tU^{\circ} = \tF^{\circ} - \I.
	\end{split}
	\label{eq:interDisplacementGradient}
\end{equation}
The gradient increment $\GradInc$ is finally found using (\ref{eq:relationBetweenDeformationGrads}), (\ref{eq:incrDisplacementGradient}) and (\ref{eq:interDisplacementGradient}) as
\begin{equation}
	\begin{split}
		\tF &= \tDeltaF \cdot \tF^{\circ} \\
		\I + \GradU &= (\I + \tGradDU) \cdot \tF^{\circ} \implies \\
		\GradU &= \tF^{\circ} + \tGradDU \cdot \tF^{\circ} - \I \implies\\
		\GradU &= \GradOld + \GradInc.
	\end{split}
	\label{eq:gradientIncrement}
\end{equation}

\bibliography{mybibfile}

\end{document}